\documentclass{aa}  
\usepackage{graphicx,natbib,url,twoopt}
\usepackage{graphicx,url,twoopt}
\usepackage[varg]{txfonts}           
\usepackage{hyperref}                

\usepackage[caption = false]{subfig}
\usepackage{graphicx}
\usepackage{longtable}
\makeatletter
\newcommand{\bibnote}[2]{\@namedef{#1note}{#2}}
\newcommand{\biblink}[2]{\@namedef{#1link}{#2}}
\makeatother

\begin{document}
\title{Photodissociation of CO in the outflow of evolved stars}
\author{{M. Saberi}
        \and W. H. T. Vlemmings
       \and  E. De Beck}

\institute{ {Department of Space, Earth and Environment, Chalmers University of Technology, Onsala Space Observatory, 43992 Onsala, Sweden \email{maryam.saberi@chalmers.se}}
}
\date{}

\abstract 
{Ultraviolet (UV) photodissociation of carbon monoxide (CO) controls the abundances and distribution of CO and its photodissociation products. This significantly influences the gas-phase chemistry in the circumstellar material around evolved stars. A better understanding of CO photodissociation in outflows also provides a more precise estimate of mass-loss rates.}
{We aim to update the CO photodissociation rate in an expanding spherical envelope assuming that the interstellar radiation field (ISRF) photons penetrate through the envelope. This will allow us to precisely estimate the CO abundance distributions in circumstellar envelope around evolved stars.}
{We used the most recent CO spectroscopic data to precisely calculate the depth dependency of the photodissociation rate of each CO dissociating line. We calculated the CO self- and mutual-shielding functions in an expanding envelope. We investigated the dependence of the CO profile on the five fundamental parameters mass-loss rate, the expansion velocity, the CO initial abundance, the CO excitation temperature, and the strength of the ISRF.}
{Our derived CO envelope size is smaller than the commonly used radius derived by Mamon et al. 1988. The difference between results varies from $1\%$ to 39$\%$ and depends on the H$_2$ and CO densities of the envelope. We list two fitting parameters for a large grid of models to estimate the CO abundance distribution. We demonstrate that the CO envelope size can differ between outflows with the same effective content of CO, but different CO abundance, mass-loss rate, and the expansion velocity as a consequence of differing amounts of shielding by H$_2$ and CO.}
{Our study is based on a large grid of models employing an updated treatment of the CO photodissociation, and in it we find that the abundance of CO close to the star and the outflow density both can have a significant effect on the size of the molecular envelope. We also demonstrate that modest variations in the ISRF can cause measurable differences in the envelope extent.}

\maketitle

\section{Introduction}\label{Introduction}

Asymptotic giant branch (AGB) stars are among the most important contributors of dust and heavy elements in the universe. These stars enrich the interstellar medium (ISM) and galaxies by ejecting a large fraction of their material through strong stellar winds. An extended circumstellar envelope (CSE) will be created around the star as a consequence of the intense mass loss \citep[e.g.][]{agb}. Understanding the complex chemical networks in the CSE of AGB stars is required for a better understanding of the enrichment and chemical evolution of the ISM and galaxies.

Carbon monoxide (CO), the most abundant molecule after molecular hydrogen (H$_2$), has been used to constrain the physical properties and chemical composition of the ISM and CSEs \citep[e.g.][]{Goldreich76, Scalo80, Millar87, Garrod06, Morata08}.
Photodissociation by ultraviolet (UV) radiation is the dominant process destroying CO and determining its abundance distribution. Therefore, a precise estimation of the CO photodissociation rate is important in both chemical and physical modelling of CSEs. 

Generally speaking, molecular photodissociation by UV radiation can dominate by direct or indirect photodissociation, depending on the molecular structure. In direct photodissociation, the photodissociation cross section is continuous as a function of photon energy (continuum photodissociation). Thus, all absorptions lead to molecular dissociation. For indirect photodissociation, the photodissociation cross section contains a series of discrete peaks (line photodissociation). Therefore, absorption at only certain wavelengths leads to molecular dissociation.
At wavelengths shorter than that of the H Lyman limit (911.7 \r{A}), CO photodissociation occurs entirely in a set of discrete UV wavelength lines. This process makes CO strongly subject to self shielding, in cases of high abundance, and to mutual shielding by other species which are dissociated at the same wavelengths such as atomic and molecular hydrogen, atomic carbon, and dust. 
The dissociating photons will be absorbed by species closer to the UV source and thus molecules at the deeper regions will be shielded. The amount of shielding depends on the UV intensity, the geometry of the cloud which determines the photon penetration probability, and the column density of species with the same dissociating wavelengths. 
\cite{Bally82} and \cite{Glassgold85} have shown the importance of the CO self-shielding in molecular clouds based on anomalous intensity ratios of various CO isotopologues which are selectively photodissociated in the edge of molecular clouds due to various column densities.

The most updated CO unshielded photodissociation rate in the interstellar medium of the solar neighborhood is estimated to be $2.6 \times 10^{-10}$ s$^{-1}$ by \cite{Visser09}. We derive the same unshielded photodissociation rate in the outflows of evolved stars assuming that the \cite{Draine78} radiation field penetrates the envelope. This rate depends only on the following: the radiation field, the accuracy of the CO spectroscopic data and the surface temperature of the astrophysical region.

However, the geometrical distances over which the photodissociation is significant depends strongly on the physical conditions of the penetrated environment such as its geometry, temperature distribution, and H$_2$ density. Thus, the environmental properties should be taken into account in calculations of the self- and mutual-shielding functions in various environment such as interstellar clouds and CSEs.

There is a good understanding of the depth dependency of the CO photodissociation mechanism in interstellar clouds and the photodissociation rate has been regularly updated with new laboratory data \citep[e.g.][]{Solomon72, Bally82, Glassgold85, vanDishoeck86, vanDishoeck88, Visser09}. The shielding functions for several sets of input parameters for interstellar clouds can be downloaded\footnote{\url{http://home.strw.leidenuniv.nl/~ewine/photo/index.php?file=CO_photodissociation.php}} from the work by \cite{Visser09}.

In case of CSEs, \citet[][hereafter MJ83]{Morris83} present the theory of calculating the depth dependency of the CO photodissociation rate by considering CO self-shielding and H$_2$ mutual-shielding in a spherical expanding envelope through a `one-band approximation'. In this approximation, they assume that CO dissociates only at 1000 \r{A}.
After higher resolution laboratory measurement of far-UV absorption and fluorescence cross sections by \cite{Letzelter87}, the rate was updated by \citet[][hereafter MGH88]{Mamon88} considering 34 dissociating-bands.
Afterwards, \cite{Visser09} collected the latest CO laboratory measurements of 855 UV dissociating lines arising in levels $J=0$ to 9 in the lowest vibrational state $v=0$.
The updated CO photodissociation rate in a CSE is presented in two recent works by \cite{Li14} and \cite{Groenewegen17}. However, both works use the shielding functions calculated for an interstellar cloud \citep{Visser09} not a CSE.
In the current work, we aim to calculate the CO photodissociation rate in a CSE using the most updated laboratory data and following the shielding functions developed for CSEs by MJ83 and MGH88. 
We discuss the differences of the depth dependency of the photodissociation rate in interstellar clouds and CSEs in Sect. \ref{The environmental dependency}.
In addition, we have investigated the effect of several more free parameters on the CO dissociation rate. 
The new treatment of the CO photodissociation has been incorporated into a chemical network describing the CSEs of AGB stars.

\section{Circumstellar chemistry}\label{Code}                              

We use an extended version of the publicly available circumstellar envelope chemical model rate13-cse code\footnote{http://udfa.ajmarkwick.net/index.php?mode=downloads} \citep{McElroy13}.
We modified calculations of the CO photodissociation rate from a one-band approximation to a treatment where all known lines are taken into consideration (see Sect. \ref{The CO ph rate}).

The code assumes a spherically symmetric envelope which is formed due to a constant mass loss $\dot{M}$. The envelope expands with a constant radial velocity $V_{\rm exp}$. The H$_2$ density falls as $1/r^2$ where $r$ measures the distance from the central star.
We used the gas temperature profile given by MGH88 which is derived for CW Leo as follows:

\begin{equation}
T_{\rm kin}(r) = 14.6 \: (\frac{r_0}{r})^{\beta} \:\:\:\: [\rm K],
\label{Temp}
\end{equation}
where $r_0$ = 9 $\times$ 10$^{16}$ cm, $\beta$ = 0.72 for $r<r_0$ and $\beta$ = 0.54 for $r>r_0$. We assume a minimum temperature of 10 K in the outer CSE.


We incorporated an extended version of the chemical network from the UMIST Database for Astrochemistry \citep{McElroy13} which includes the $^{13}$C and $^{18}$O isotopes, all corresponding isotopologues, their chemical reactions and the properly scaled reaction rate coefficients \citep{Rollig13}.
The chemical network includes 933 species and 15108 gas-phase reactions. The isotopologue chemistry will be discussed in detail in a forthcoming paper.

\subsection{CO spectroscopic data} \label{The CO spectroscopic data} 

The dissociation energy of the CO ground state is 11.09 eV, thus the photodissociation occurs in the wavelength range 911.75 - 1117.8 \r{A}. 
The CO photodissociation occurs entirely through line absorptions into pre-dissociating states \citep[e.g.][]{Visser09}. 
We use the latest laboratory measurements of CO data \citep{Visser09} which includes 855 UV transitions containing rotational excitations $J=0-9$ to calculate the photodissociation rate. 
The higher excitation levels up to $J=30$ were examined in the calculations and since the effect was marginal, we excluded the higher transitions to increase the time efficiency of the code.

\subsection{CO photodissociation rate}\label{The CO ph rate}

The total CO photodissociation rate at radius $r$ is the summation of the photodissociation rates of all discrete contributing lines $i$ as follows:
\begin{equation}
 k(r) = \sum_{i=1}^{855} k_i^0 \: \beta_i(r) \: \gamma_i(r) \:\:\:\:  [s^{-1}],
\label{K(r)}
\end{equation}
where $k_i^0$ is the CO unshielded photodissociation rate at the edge of the cloud, $\beta_i$ counts the CO self-shielding efficiency and $\gamma_i$ counts the mutual shielding from other species. 
$k_i^0$ is estimated to be $2.6 \times 10^{-10}$ s$^{-1}$ and the detailed calculations are presented in the Appendix \ref{The unshielded ph rate}.
We note that in calculations of the photodissociation rates we assumed the CO excitation temperature $T_{\rm ex}$ to be the same as the gas kinetic temperature $T_{\rm kin}$ given in Eq. \ref{Temp}.

Dissociating radiation can be absorbed by CO (self-shielding), H, C, H$_2$ and dust (mutual-shielding) \citep[e.g. MGH88,][]{Visser09}.
The mutual shielding by different species depends on their column density and the amount of line overlaps in the relevant region of the spectrum. 
The amount of mutual shielding by dust is assumed to depend on the total number of protons [$n(\rm H)$+$2n(\rm H_2)$] which determine the dust extinction.
In our models, the column densities of C and H are insufficient to produce very much blocking. 
Thus, the shielding by dust and H$_2$ dominates the CO mutual shielding.
We note that in the presence of extra UV radiation from a hot binary companion and/or stellar chromospheric activity there could be an enhancement of the atomic abundances.
The investigation of how and whether these would impact the CO abundance distribution is, however, beyond the scope of the current study.

Calculations of the depth dependency of the CO photodissociation rate require accurate information on the line wavelength, oscillator strengths, the pre-dissociation probabilities and the line widths of CO and H$_2$. 
We used the new compiled H$_2$ line list (J. Black, private communication) based on energy levels and transition probabilities computed by \cite{Abgrall94, Abgrall97}.
In the Appendix \ref{COshieldingfunctions}, we review the underlying physics of calculations of shielding functions $\beta$ and $\gamma$ in an expanding CSE.

\begin{figure}[t]
  \centering
  \includegraphics[width=85mm]{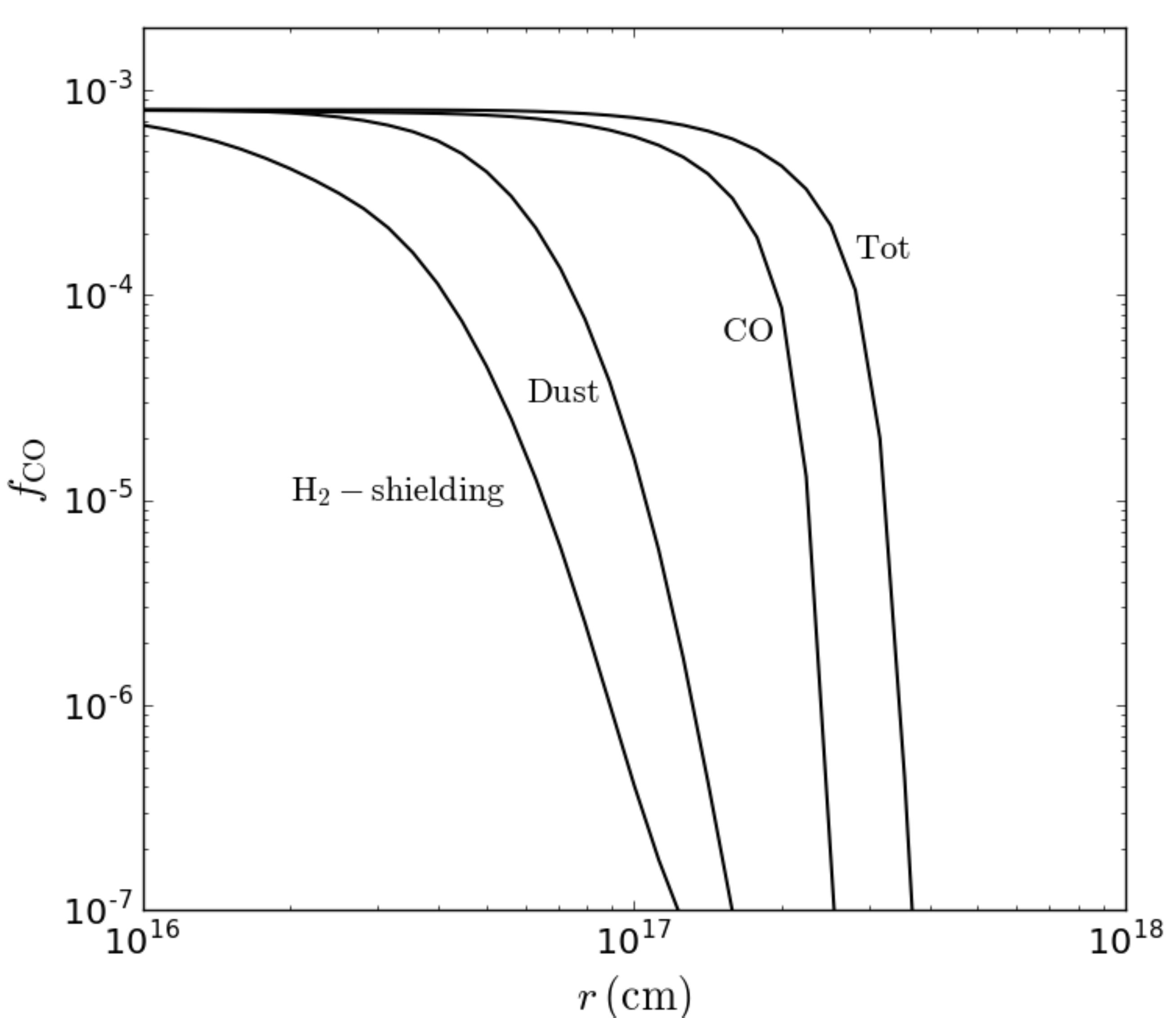}
  \caption[]{\label{SH}
     CO fractional abundance distributions for simulations with different shielding functions that regulate CO photodissociation: shielding by H$_2$, shielding by dust, CO self-shielding, and the total shielding for the reference model.}
\label{SH}  
\end{figure}
\begin{figure}[t]
  \centering
  \includegraphics[width=82mm]{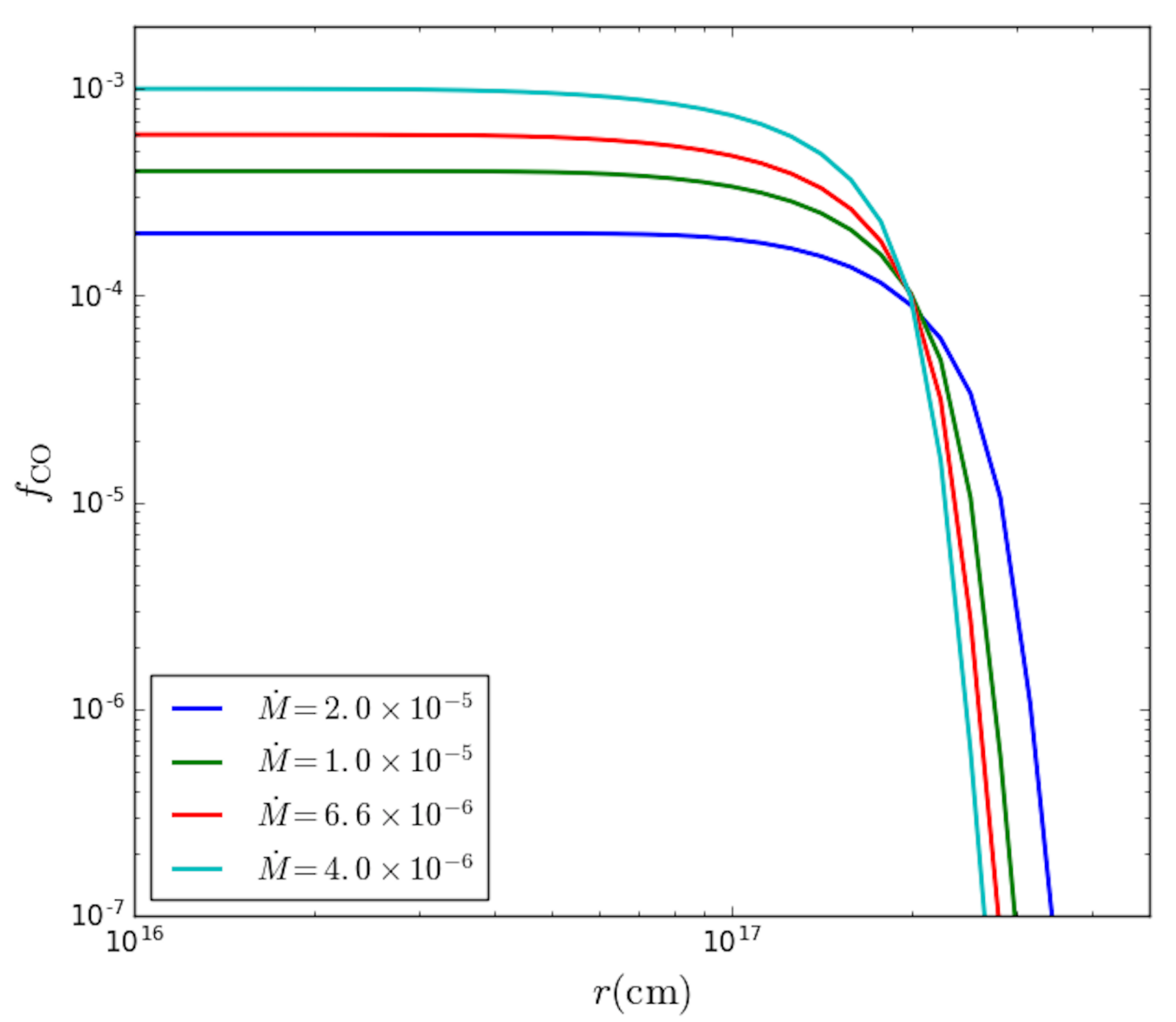}
  \caption[]{\label{fCO}
    CO fractional abundance distributions for models with a constant expansion velocity $V_{\rm exp} = 15$ [km s$^{-1}$] and various $\dot{M} \:[M_{\odot} \rm yr^{-1}]$ and $f_{\rm CO}$ in a way to keep the same amount of effective CO for all models.}
\label{MLfCO1}  
\end{figure}

\begin{figure}[t]
  \centering
  \includegraphics[width=85mm]{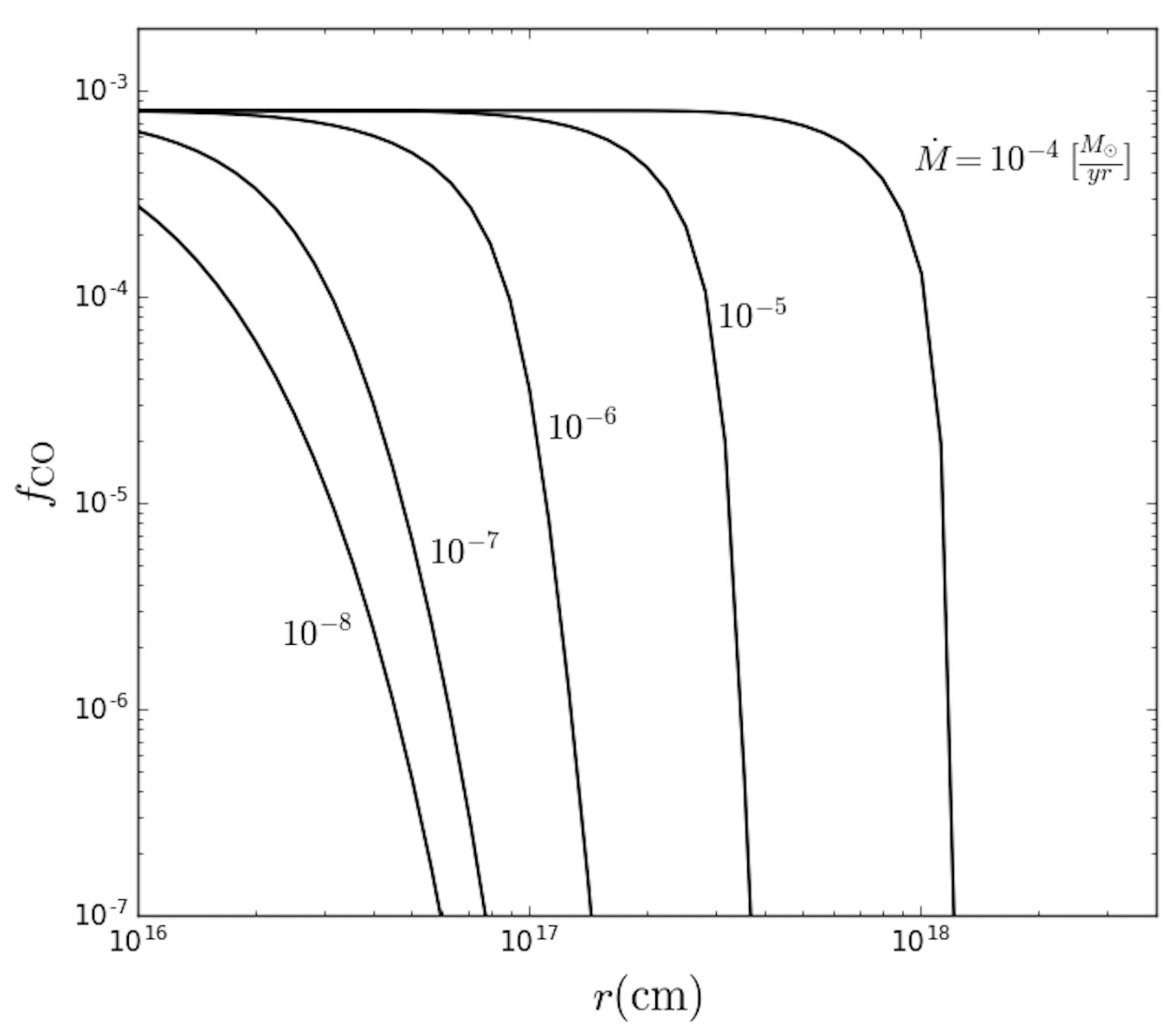}
  \caption[]{\label{ML}
     CO fractional abundance distributions for models with $f_{\rm CO} = 8 \times 10^{-4}$, $V_{\rm exp}=15$ [km s$^{-1}$], and a range of mass-loss rates which are marked in the figure.}
\label{ML}  
\end{figure}


\begin{figure}[t]
  \centering
  \includegraphics[width=85mm]{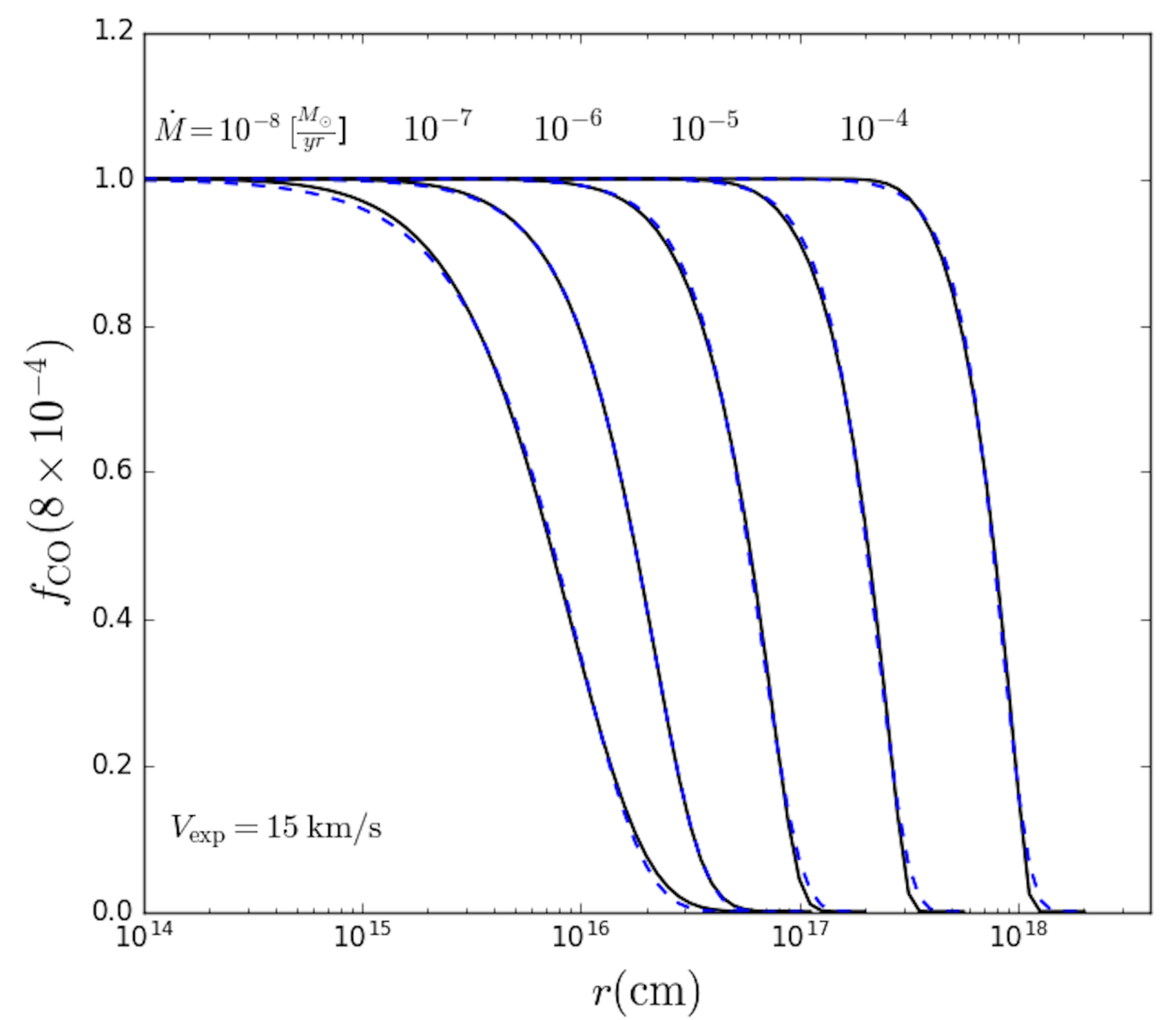}
  \caption[]{\label{Fit}
  Comparison of the CO fractional abundance distributions from modelling (solid black lines) and the analytic fitting formula (dashed blue lines) for models with $f_{\rm CO} = 8 \times 10^{-4}$, $V_{\rm exp}$ = 15 [km s$^{-1}$] and a range of $10^{-8} <\dot{M}< 10^{-4} \, [M_\odot \rm yr^{-1}]$.}
\label{Fit}  
\end{figure}

\begin{table}[t]
  \centering
  \setlength{\tabcolsep}{5.0pt}
    \caption{\tiny Envelope parameters and assumptions for the reference model.}
  \begin{tabular}{@{} cccccc@{}}
\hline
$\dot{M}$ & $V_{\rm exp}$ & $r_{\rm in}$ & $T_{\rm kin}$ & $f_{\rm CO}^1$ \\
\hline
\:
$[M\odot \rm yr^{-1}]$ & [km s$^{-1}$] & [cm] & [K] & \\
\hline
 $1\times10^{-5}$ & 15 & $1\times10^{14}$ & Eq.\ref{Temp} & $8\times10^{-4}$ &
 \end{tabular}
\tablefoot{1. $f_{\rm CO}$ is the initial abundance of CO relative to H$_2$.}
 \label{SModel}
\end{table}

\section{Results}\label{Results}

To study the significance of the different shielding processes we use a reference model with the physical parameters listed in Table \ref{SModel}. This model serves as a direct comparison to the work by MGH88.
Figure \ref{SH} shows the fractional CO abundance profile calculated by considering different shielding contributions from CO, H$_2$, dust, and the total shielding for the reference model. As we can see, the CO self-shielding plays a major role in the total shielding.

In Fig. \ref{MLfCO1} we illustrate how the CO abundance distribution varies for models with different initial CO abundances $f_{0}$ and mass-loss rates $\dot{M}$ while preserving the total amount of CO ejected by the star. It is clear that the H$_2$ density (set by $\dot{M}$ and $V_{\rm exp}$) dictates the size of the envelope, while the initial CO abundance affects the steepness of the slope.
This differentiation provides a way to break the degeneracy between $f_{\rm CO}$ and $\dot{M}$ encountered when both low-$J$ and high-$J$ CO transitions are used in CO radiative transfer (RT) modelling. 

Figure \ref{ML} presents the variation of the CO distribution profiles with $\dot{M}$, keeping all other parameters the same as the reference model. 
An increase in $\dot{M}$ translates into a stronger shielding and hence a larger CO envelope with a sharper drop-off at the outer edge.

The CO abundance profiles derived from our simulations can be fitted by the analytical formula derived by MGH88 as follows:
\begin{equation}
 f_{\rm CO} = f_0 \: \rm{exp\big(- ln(2) \: (\frac{r}{ r_{1/2}})^{\alpha}\big)},
 \label{f}
\end{equation}
where $f_0$ is the initial CO abundance, $\alpha$ determines the steepness of the profile and $r_{1/2}$ marks the radius where the CO abundance drops to half of its initial value. 
Figure \ref{Fit} shows the accuracy of the fitting formula for a range of mass-loss rates; all other parameters are the same as the reference model given in Table \ref{SModel}. 
In general, $f_0$ depends on the chemical type of AGB star and is commonly assumed to have average values $(2-6-10)\times 10^{-4}$ for M-, S-, and C-type AGB stars, respectively \citep{Ramstedt14}.

\begin{figure*}
\center
\subfloat{\includegraphics[width = 3.5in]{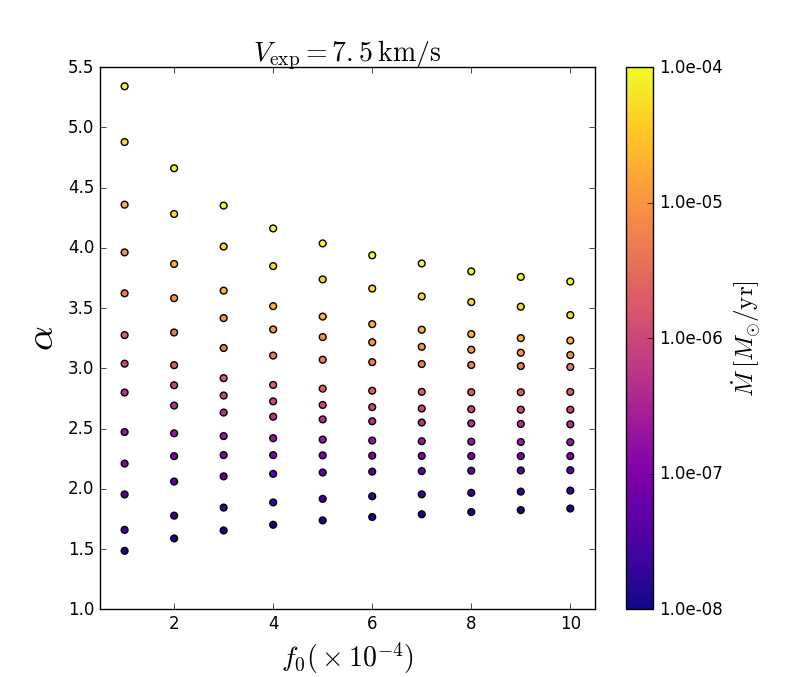}} 
\subfloat{\includegraphics[width = 3.5in]{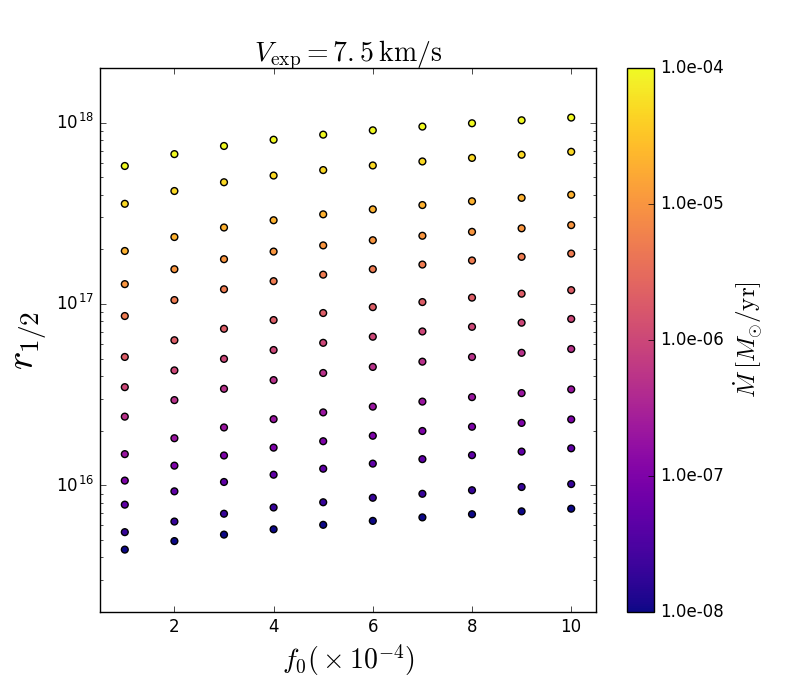}}\\
\subfloat{\includegraphics[width = 3.5in]{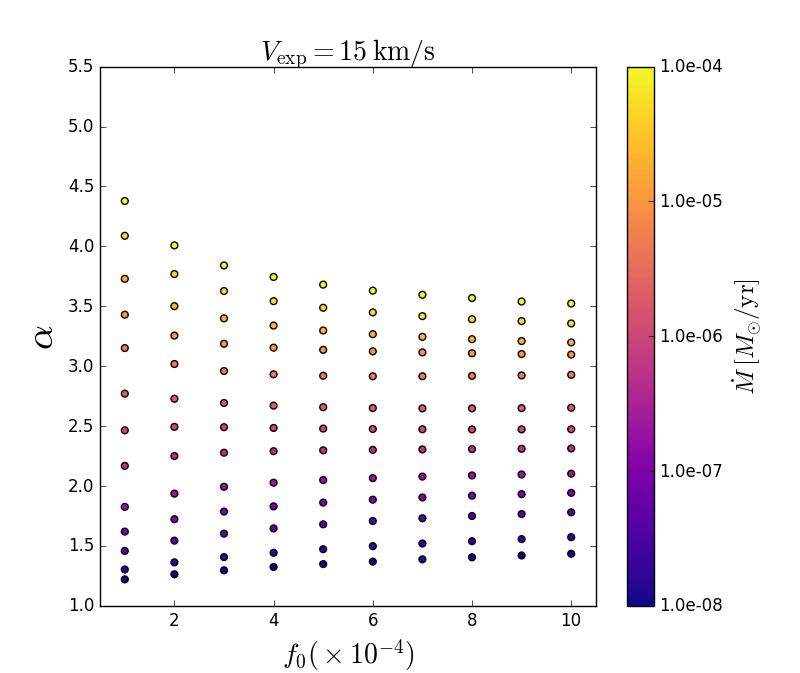}}
\subfloat{\includegraphics[width = 3.5in]{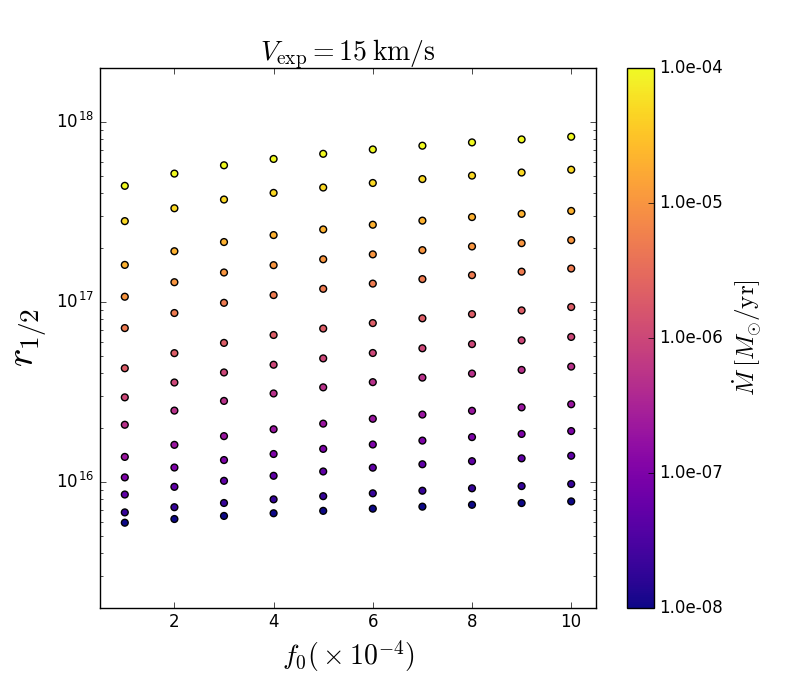}}\\
\subfloat{\includegraphics[width = 3.5in]{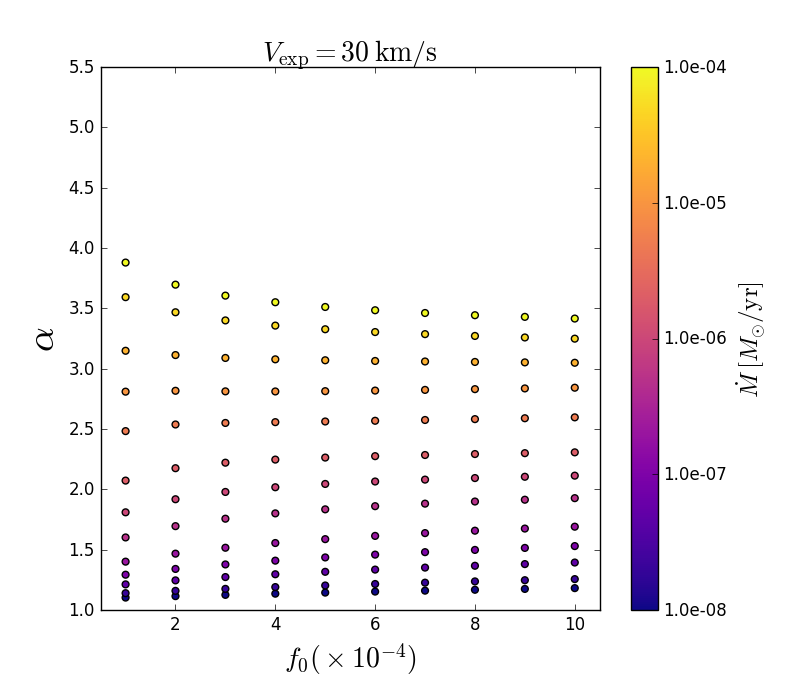}}
\subfloat{\includegraphics[width = 3.5in]{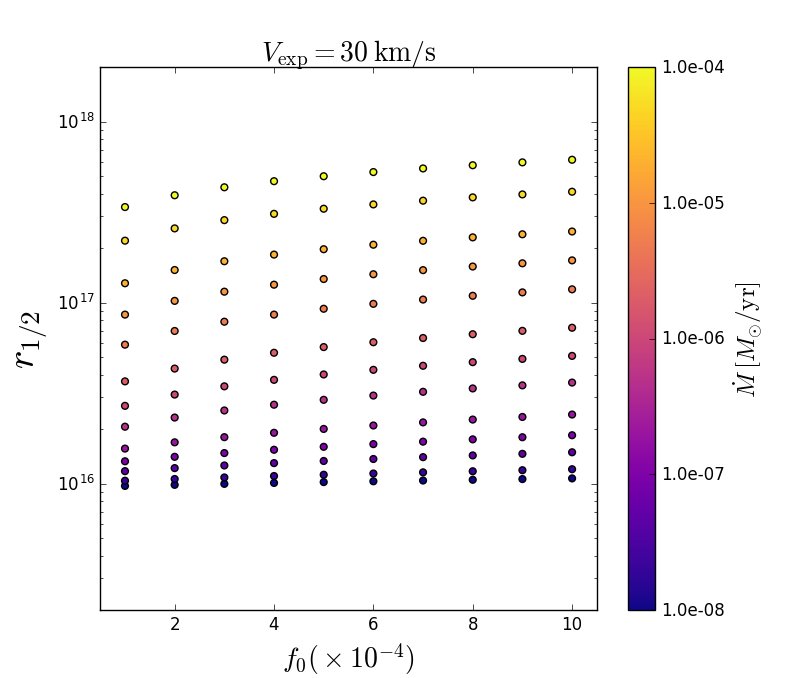}} 
\caption{Variation of two fitting parameters $\alpha$ and $r_{1/2}$ for all models presented in Table \ref{FullG}.}
\label{AllFitP}
\end{figure*}

We derive $\alpha$ and $r_{1/2}$ for a grid containing 390 models with varying $f_{0}$, $\dot{M}$, and $V_{\rm exp}$.
We considered ten values for $f_0$ ($1,2,\cdots,10\times10^{-4}$), 13 values for $\dot{M}$ ($[1,2,5]\times10^{-8}, [1,2,5] \times 10^{-7}, \cdots, 1\times10^{-4}$\, [$M_{\odot}$ yr$^{-1}$]), and three values for $V_{\rm exp}$ (7.5, 15, 30 [km\,s$^{-1}$]). 
Table \ref{FullG} lists the resulting $\alpha$ and $r_{1/2}$ values for all models. 
Figure \ref{AllFitP} shows how variations of $f_{\rm CO}$, $\dot{M}$, and $V_{\rm exp}$ affect $\alpha$ and $r_{1/2}$.
Increasing $\dot{M}$ and decreasing $V_{\rm exp}$ leads to an enhancement of the $n_{\rm H_2}$ and thus more shielding and a larger CO envelope. Similarly an increase in $f_{\rm CO}$ enhances the CO self-shielding. 
The most drastic changes in $\alpha$ and $r_{1/2}$ happen for oxygen-type AGB stars with lower CO abundance, where the CO self-shielding is not very efficient.

\section{Discussion}\label{Discussion}

\subsection{Influence of the temperature on the CO abundance distribution}\label{TempCO}

We examined the influence of the gas kinetic temperature profile on the CO envelope size for the reference model. The CO excitation temperature is assumed to be the same as the gas kinetic temperature profile which is given in Eq. \ref{Temp}.
We assumed the temperature and radius at the inner envelope to be $T_0=2000$ K and $r_0=10^{14}$ cm.
We considered $0.4<\beta<1.0$ which reasonably covers the gas temperature profile of AGB CSEs
\citep{DeBeck12, Danilovich14, Khouri14, Maercker16, Ramos18, VandeSande18}. $\beta$ determines the slope of the profile.
Figure \ref{TempCO} shows the considered temperature profiles and their corresponding CO abundance distributions. 
The fitting parameters vary in the ranges $1.65\times 10^{17} <r_{1/2}< 2.03\times10^{17}$ cm and $3.00 <\alpha< 3.24$.
The temperature profiles with $\beta \geq 0.6$ give rise to similar CO abundance distributions whereas those with $\beta<0.6$ lead to smaller envelopes. This is due to a lower shielding effectiveness in the hotter envelope (see Appendix \ref{COshieldingfunctions}). 

We also examined the influence of constant CO excitation temperature $T_{\rm ex}=5 , 20, 50, 100$ K as assumed by \cite{Groenewegen17} versus $T_{\rm ex}=T_{\rm kin}$ for the reference model.
The two fitting parameters vary in ranges $1.53\times 10^{17} <r_{1/2}< 2.25\times10^{17}$ cm and $2.88 <\alpha< 3.41$. As shown in Fig. \ref{COTex} a lower CO excitation temperature results in a bigger CO envelope.
The reduced $T_{\rm ex}$ leads to higher lower-level populations $x_l$ (see Appendix \ref{COshieldingfunctions}) and thus more efficient shielding and therefore a bigger envelope.

\begin{figure*}
\centering
\begin{minipage}{.5\textwidth}
  \centering
  \includegraphics[width=72mm]{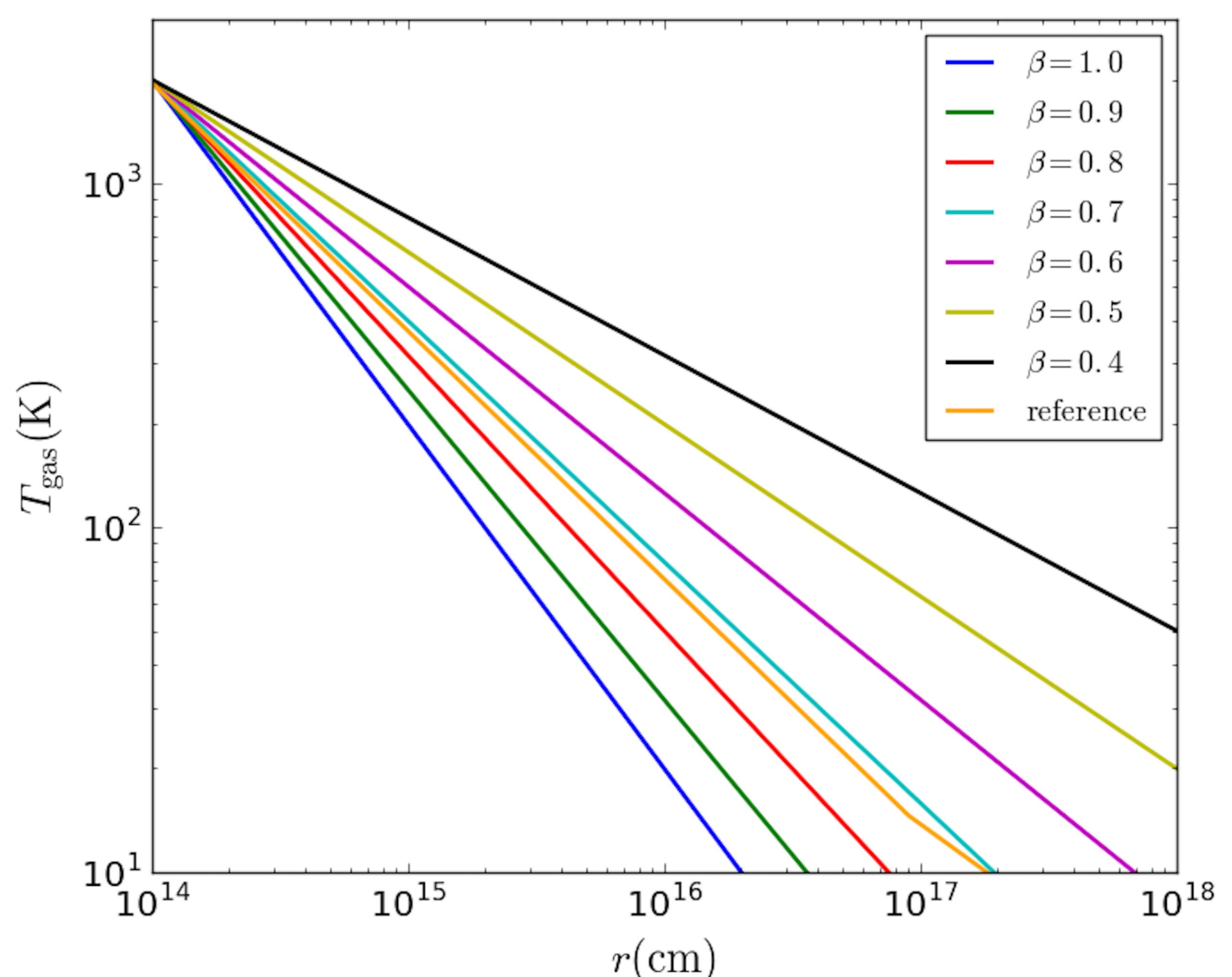}    
\end{minipage}%
\begin{minipage}{.5\textwidth}
  \centering
  \includegraphics[width=68mm]{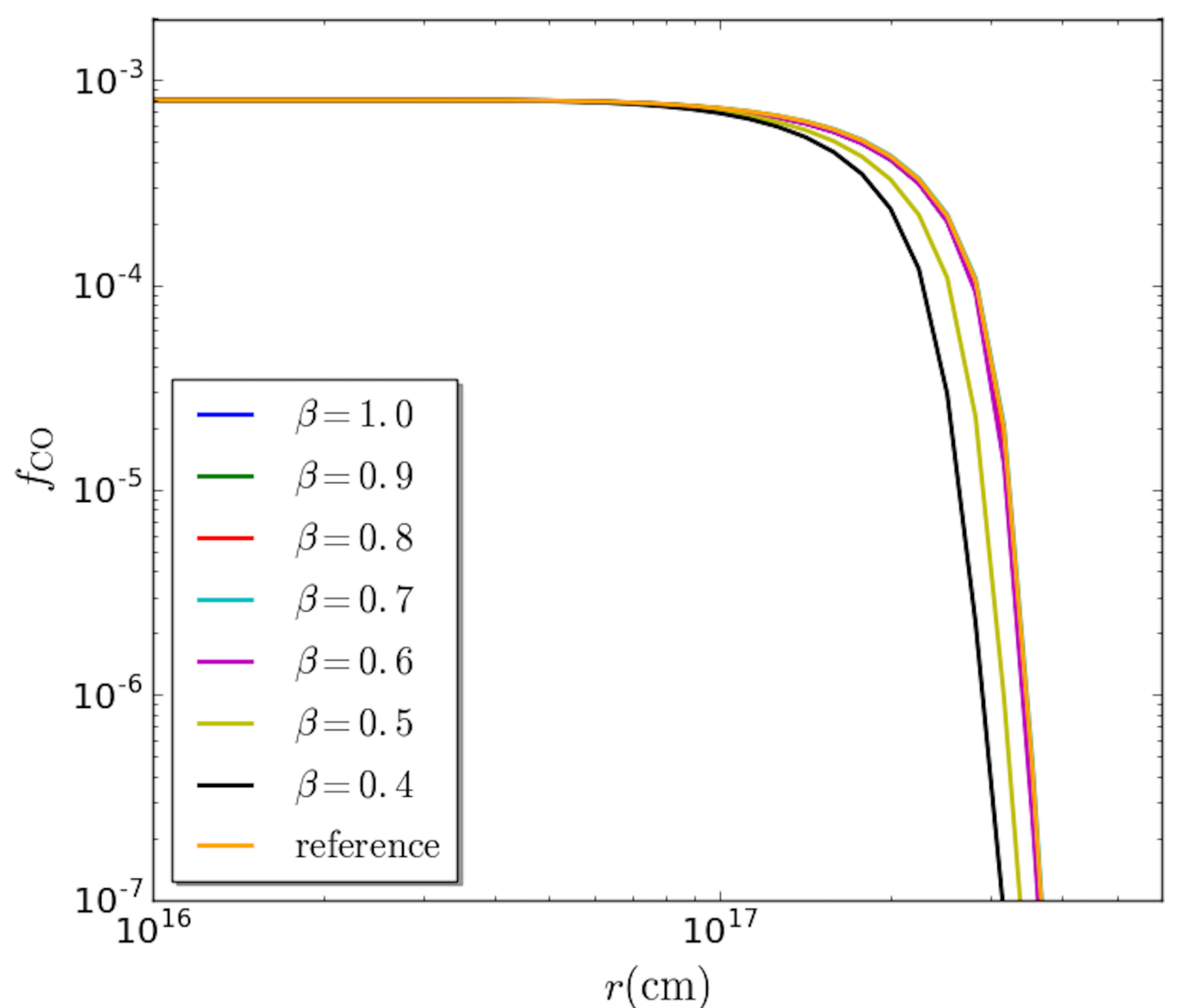}   
\end{minipage}
\caption[]{\label{TempCO} 
Left panel: Tested gas kinetic temperature profiles with different $\beta$ values. The profile with label 'reference' corresponds to the temperature of the reference model. Right panel: CO abundance distribution profile derived using the temperature profiles from the left panel.}
\end{figure*}

\begin{figure}
  \centering
  \includegraphics[width=85mm]{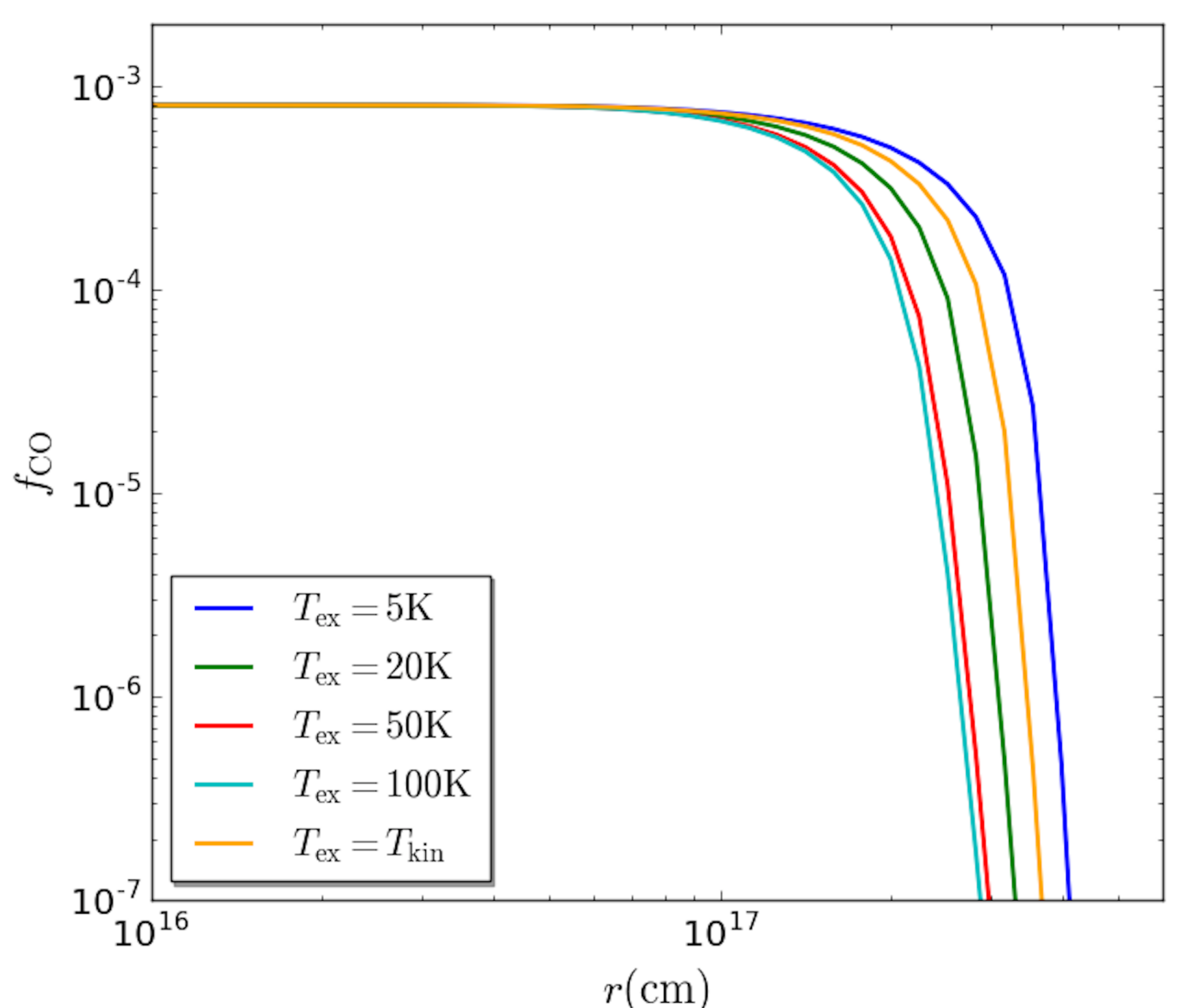}
  \caption[]{\label{COTex}
    CO fractional abundances distributions for the reference model when different CO excitation temperature profiles are considered in calculations of the CO photodissociation rate.}
\label{COTex}  
\end{figure}

\subsection{Influence of the ISRF on the CO abundance distribution}\label{ISRF}

The Draine ISRF that we considered in this work has been measured in the solar neighborhood. \cite{McDonaldZ15} show that the ISRF is considerably higher in globular clusters. 
This will have a significant impact on the size of the CO envelope of evolved stars which are located in clusters \citep{McDonald15}.
On the other hand, objects which lie above the Galactic plane, for example IRC+10216, are possibly exposed to a weaker ISRF.
We investigated the influence of the strength of the ISRF on the CO envelope size. We scaled the \cite{Draine78} radiation field that we used for the reference model by factors $\chi=0.25, 0.5, 1, 2,$ and 4. Table \ref{ISRFT} presents the results for models with the mass-loss rates $\dot{M}= 10^{-8}, 10^{-7}, 10^{-6}, 10^{-5}, 10^{-4}$ [$M_{\sun} \rm yr^{-1}$]. We considered the expansion velocity $V_{\rm exp}= 15$ [km s$^{-1}$] and the initial CO abundance of $f_{\rm CO} = 8 \times 10^{-8}$ for all models. As expected, the effect of the ISRF strength on the CO envelope size is more prominent in stars with low mass-loss rates with weaker shielding efficiency.  
Increasing the ISRF by a factor of two reduces the CO envelope size by $40, 34, 28, 20, 14 \%$ for stars with $\dot{M}=10^{-8}, 10^{-7},10^{-6},10^{-5},10^{-4}$  \: $[M_{\odot} \: \rm yr^{-1}$], respectively. 

We note that in some cases, there is extra UV radiation which internally penetrates into the CSE. The inner UV radiation can arises from stellar chromospheric activity and/or a hot binary companion \citep[e.g.][]{Montez17}. In hot post-AGB stars, the high temperature of the star itself also generates UV photons. 
In such cases, the inner UV radiation will be quickly absorbed by inner dust and thus is not expected to affect the extent of the CO envelope. However this likely affects the CSE chemistry \citep[e.g.][]{Saberi17, Saberi18, VandeSande19} in the dust formation region, which is beyond the scope of this paper.

\begin{table*}
 \centering
  \setlength{\tabcolsep}{4.0pt}
\caption{Fitting parameters of the CO envelope size for a range of ISRF intensity. The CO abundance and the expansion velocity are assumed to be $f_{\rm CO/H_2} = 8 \times 10^{-4}$ and $V_{\rm exp}= 15 \:[\rm km \: s^{-1}$] for all models. $\chi$ represents the ISRF scaling factor.}
\begin{tabular}{c|cc|cc|cc|cc|cc}
\hline
 &  $\chi =0.25$   && $\chi =0.5$    && $\chi =1$   && $\chi =2$  && $\chi =4$ \\
\hline
 $\dot{M} [M_{\sun} \rm yr^{-1}]$  & $\alpha$  & $r_{1/2}$ [cm]  & $\alpha$  & $r_{1/2}$ [cm] & $\alpha$  &  $r_{1/2}$ [cm] & $\alpha$ &  $r_{1/2}$ [cm] & $\alpha$ &  $r_{1/2}$ [cm]\\                
\hline
$1\times10^{-8}$ & 1.23 & $2.28\times10^{16}$ & 1.30 & $1.28\times10^{16}$ & 1.40 & $7.44\times10^{15}$ & 1.51 & $4.43\times10^{15}$ & 1.64 & $2.72\times10^{15}$ \\
$1\times10^{-7}$& 1.61 & $4.40\times10^{16}$ & 1.76 & $2.75\times10^{16}$ & 1.91 & $1.76\times10^{16}$ & 2.08 & $1.16\times10^{16}$ &  2.26 & $7.79\times10^{15}$ \\
$1\times10^{-6}$&  2.01 & $1.16\times10^{17}$ &2.23 & $8.21\times10^{16}$ &  2.47 & $5.81\times10^{16}$ & 2.71 & $4.15\times10^{16}$ &  2.95 & $3.01\times10^{16}$\\
$1\times10^{-5}$& 2.26 & $3.20\times10^{17}$ &  2.67 & $2.54\times10^{17}$ & 3.10 & $2.02\times10^{17}$ &3.52 & $1.62\times10^{17}$ &  3.91 & $1.30\times10^{17}$ \\
$1\times10^{-4}$ & 2.56 & $9.99\times10^{17}$ &3.04 & $8.83\times10^{17}$&  3.56 & $7.67\times10^{17}$ & 4.12 & $6.59\times10^{17}$ &  4.67 & $5.63\times10^{17}$ \\
\hline
\end{tabular}
\label{ISRFT}
\end{table*}

\subsection{{Environmental dependency of the CO photodissociation rate}} \label{The environmental dependency}

The differences in the physical and chemical properties between interstellar clouds and CSEs can affect the CO shielding functions and thus the depth dependency of CO photodissociation rates. Here we list the environmental differences between interstellar clouds \citep{Visser09} and CSEs (this work) that enter in calculations of shielding functions: 
\begin{itemize}

\item[(a)] Model geometry: 
In interstellar cloud models, plane-parallel geometry has been considered while in the CSE model spherically symmetric geometry is assumed.
The difference in the geometry affects the photon penetration probability.  

\item[(b)] CO excitation temperature:
A constant temperature that can be chosen among the values of 5 - 20 - 50 -100 K is available in interstellar clouds model.
In the modelling of the CSE, we consider $T_{\rm ex}$ (CO) to be the same as $T_{\rm kin}$ which varies in a range $\sim10 - 2000$ K. This affects the fractional population of the lower-level especially in the inner CSE. 
Assuming a too low $T_{\rm ex}$ leads to an overestimation of the CO self-shielding as shown in Fig. \ref{COTex}.
\item[(c)] Atomic and molecular line broadening:
The Doppler and natural broadenings are the dominant processes which control the CO, H$_2$ and H line widths in the interstellar clouds.
Since the CSE around AGB stars is expanding at velocities of typically a few up to some 30 km s$^{-1}$, the expansion velocity should also be considered in the calculations of the line widths. 
\end{itemize}

\subsection{Comparison with literature}\label{Compliterature}

\begin{figure*}
\centering
\begin{minipage}{.5\textwidth}
  \centering
  \includegraphics[width=70mm]{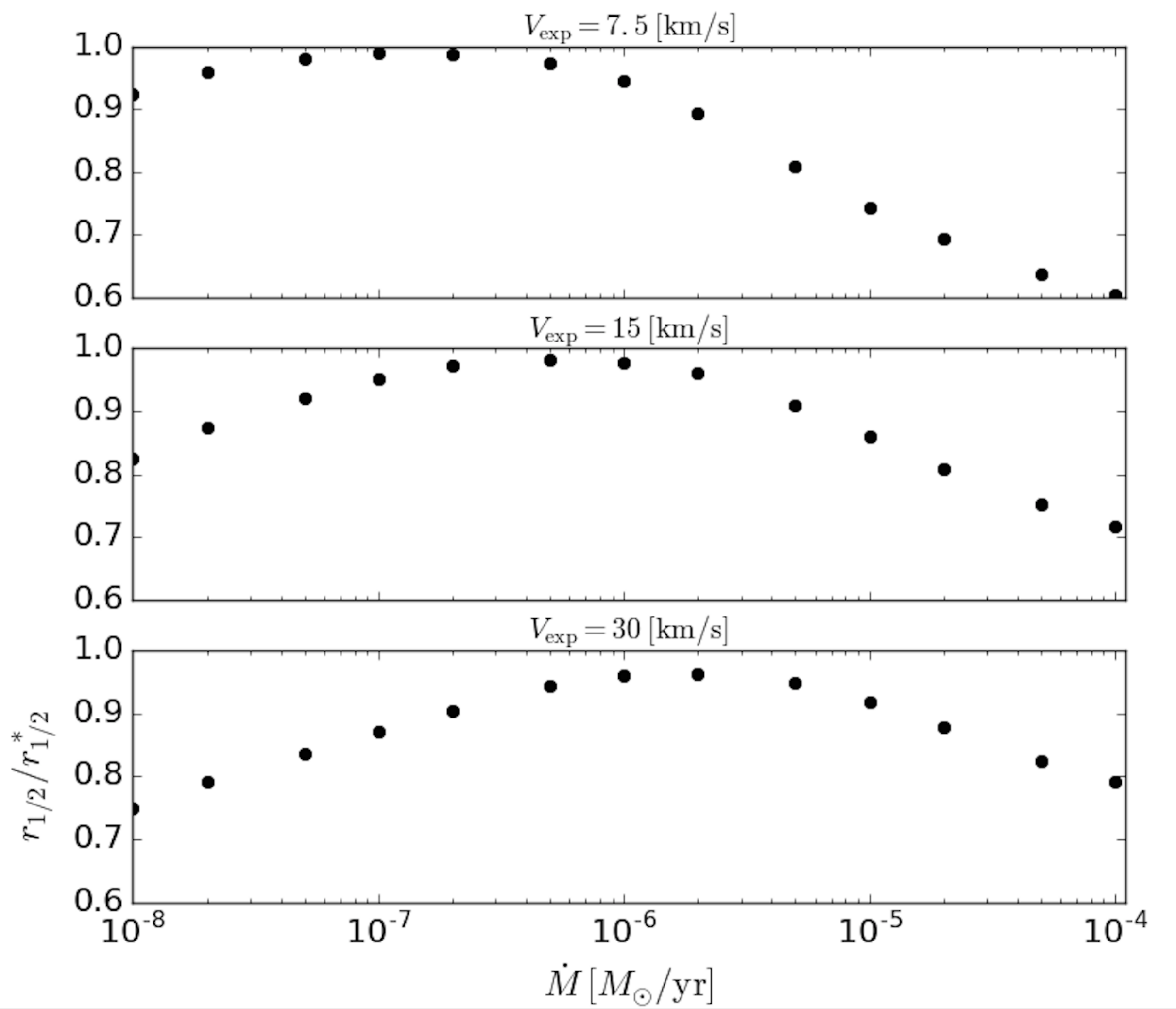} 
\end{minipage}%
\begin{minipage}{.5\textwidth}
  \centering
  \includegraphics[width=70mm]{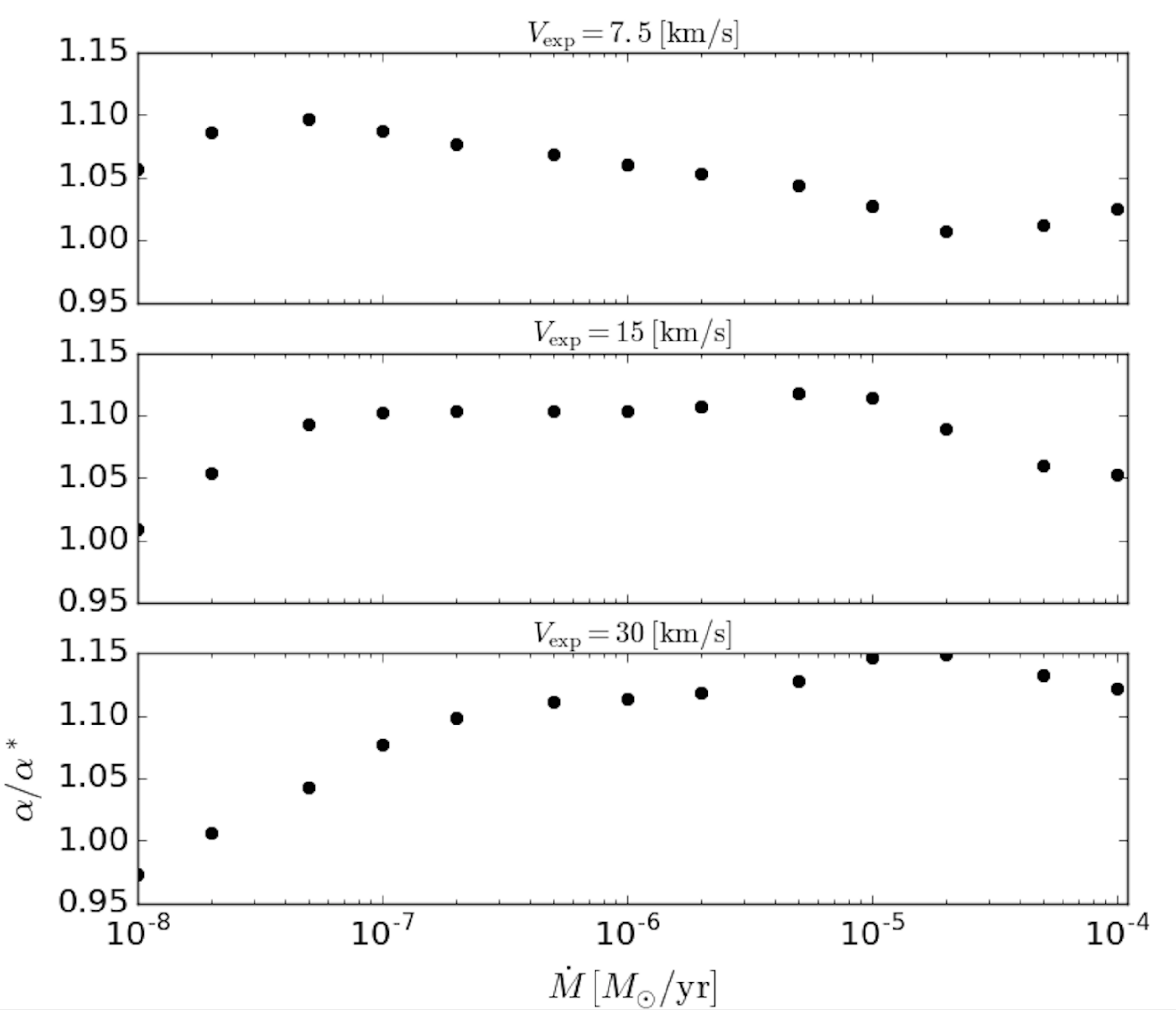}
\end{minipage}
\caption{\label{Fitpara} 
Comparison of CO abundance distribution parameters to MGH88 results. The left panel shows $r_{1/2}/r^{*}_{1/2}$ ratio and the right panel shows $\alpha/\alpha^{*}$ with $r_{1/2}^{*}$ and $\alpha^{*}$ are the MGH88 values.
}
\end{figure*}

The most commonly used method to derive the size of the CO envelope is the one presented by MGH88. 
They used the \cite{Jura74} radiation field to calculate the CO photodissociation rate. The Jura radiation field in addition to more UV observations at longer wavelengths up to 2000 \r{A} has been later used by \cite{Draine78} to derive an analytical formula for the standard ISRF \citep{Lee84}.
 Thus, in principle in the wavelength range 930-1125 \r{A}, both Draine and Jura radiation fields represent the same UV observational data.
MGH88 presented the fitting parameters of the CO envelope size for a grid of models with a constant CO abundance of $f_{\rm CO/H_2} = 8 \times 10^{-8}$ and 13 varying mass-loss rates and three expansion velocities.
 However, in addition to now outdated low-resolution CO laboratory data which causes underestimation of the unshielded photodissociation rate by $30\%$, a major drawback of this work is that it is based on calculations that assume one fixed value of $f_0=8 \times 10^{-4}$. We clearly demonstrate in the previous section that the role of $f_0$ is non-negligible in the overall dissociation efficiency. 
Table \ref{Fittable} and Fig. \ref{Fitpara} compare the MGH88 results for the $r_{1/2}$ and $\alpha$ fitting parameters with those derived in this work. 
The difference in fitting parameters ranges from 0.6-15$\%$ for $\alpha$ and 1-39$\%$ for $r_{1/2}$ between two works.
In almost all tested cases, our models predict the steepness parameter $\alpha$ to be larger than that derived by MGH88, indicating a less efficient shielding of CO in our study. In line with this, we calculate consistently smaller values for $r_{1/2}$. 
This is consistent with observational data for W Hya \citep{Khouri14}, TX Cam \citep{Ramstedt08}, and R Dor \citep{Maercker16} for example. Models of these objects reproduce the observed line emission better when they assume a smaller CO envelope than is predicted by MGH88, in line with the results of this paper.

Our derived CO radius is also smaller by $11-60\%$ compared to \cite{Groenewegen17} depending on the H$_2$ density of the envelope. This comes from different shielding functions used to calculate the depth-dependent CO photodissociation rate.

\begin{table}
  \centering
  \setlength{\tabcolsep}{3.5pt}
\caption{Comparison of the fitting parameters of the CO envelope between this work and MGH88. The CO abundance is assumed to be $f_{\rm CO/H_2} = 8 \times 10^{-4}$ for all models.}
\begin{tabular}{c|cc|cc}
\hline
\multicolumn{5}{c}{$V_{\rm exp}=7.5$ km s$^{-1}$} \\
 \hline
& This work  & &  MGH88 & \\
\hline
$ \dot{M} [M_{\sun} \rm yr^{-1}]$    &   $r_{1/2}$ [cm] & $\alpha$   &     $r_{1/2}$ [cm] &  $\alpha$    \\                
\hline
$1\times10^{-8}$ & $6.92\times10^{15}$ & 1.80 & $7.50\times10^{15}$ & 1.71 \\ 
$2\times10^{-8}$& $9.38\times10^{15}$ & 1.96 & $9.79\times10^{15}$ & 1.81  \\
$5\times10^{-8}$ & $1.46\times10^{16}$ & 2.14 & $1.49\times10^{16}$ & 1.96  \\
$1\times10^{-7}$ & $2.10\times10^{16}$ & 2.27 & $2.12\times10^{16}$ & 2.09\\
$2\times10^{-7}$ & $3.06\times10^{16}$ & 2.39 &  $3.10\times10^{16}$ & 2.22  \\
$5\times10^{-7}$ & $5.09\times10^{16}$ & 2.54 & $5.23\times10^{16}$ & 2.38  \\
$1\times10^{-6}$ & $7.47\times10^{16}$ & 2.66 & $7.91\times10^{16}$ & 2.51 \\
$2\times10^{-6}$ & $1.08\times10^{17}$ & 2.80 & $1.21\times10^{17}$ & 2.66  \\
$5\times10^{-6}$ & $1.73\times10^{17}$ & 3.02 & $2.14\times10^{17}$ & 2.90 \\
$1\times10^{-5}$ & $2.49\times10^{17}$ & 3.15 & $3.35\times10^{17}$  & 3.07 \\
$2\times10^{-5}$ & $3.68\times10^{17}$ & 3.28 & $5.31\times10^{17}$& 3.26  \\
$5\times10^{-5}$ & $6.38\times10^{17}$ & 3.55 & $9.99\times10^{17}$ & 3.51  \\
$1\times10^{-4}$& $9.91\times10^{17}$ & 3.80 & $1.64\times10^{18}$ & 3.71 \\
\hline
\hline
\multicolumn{5}{c}{$V_{\rm exp}=15$ km s$^{-1}$} \\
 \hline
  \hline
$1\times10^{-8}$ & $7.44\times10^{15}$ & 1.40 & $9.01\times10^{15}$  & 1.39 \\ 
$2\times10^{-8}$ & $9.18\times10^{15}$  & 1.53 & $1.05\times10^{16}$  & 1.46  \\
$5\times10^{-8}$& $1.29\times10^{16}$ & 1.74 & $1.40\times10^{16}$ & 1.60  \\
$1\times10^{-7}$ & $1.76\times10^{16}$ & 1.91 & $1.85\times10^{16}$ & 1.74\\
$2\times10^{-7}$ & $2.47\times10^{16}$ & 2.08 & $2.54\times10^{16}$ & 1.89  \\
$5\times10^{-7}$ & $3.98\times10^{16}$ & 2.30 & $4.05\times10^{16}$ & 2.09  \\
$1\times10^{-6}$ & $5.81\times10^{16}$ & 2.47 & $5.95\times10^{16}$& 2.24 \\
$2\times10^{-6}$ & $8.52\times10^{16}$ & 2.64 & $8.88\times10^{16}$ & 2.39  \\
$5\times10^{-6}$ & $1.40\times10^{17}$ & 2.91 & $1.54\times10^{17}$ & 2.61  \\
$1\times10^{-5}$ & $2.02\times10^{17}$ & 3.10 & $2.35\times10^{17}$ & 2.79 \\
$2\times10^{-5}$ & $2.95\times10^{17}$ & 3.22 & $3.65\times10^{17}$ & 2.96  \\
$5\times10^{-5}$ & $5.01\times10^{17}$ & 3.39 & $6.67\times10^{17}$& 3.20  \\
$1\times10^{-4}$ & $7.67\times10^{17}$ & 3.56 & $1.07\times10^{18}$ & 3.39  \\

\hline 
\hline
\multicolumn{5}{c}{$V_{\rm exp}=30$ km s$^{-1}$} \\
 \hline
  \hline
$1\times10^{-8}$ & $1.04\times10^{16}$ & 1.16 & $1.39\times10^{16}$ & 1.20 \\ 
$2\times10^{-8}$ & $1.17\times10^{16}$ & 1.23 & $1.48\times10^{16}$ & 1.23  \\
$5\times10^{-8}$ & $1.43\times10^{16}$ & 1.36 & $1.71\times10^{16}$&1.31  \\
$1\times10^{-7}$ & $1.75\times10^{16}$ & 1.49 & $2.01\times10^{16}$ & 1.39 \\
$2\times10^{-7}$ & $2.25\times10^{16}$ & 1.65 & $2.49\times10^{16}$ & 1.51  \\
$5\times10^{-7}$ & $3.35\times10^{16}$ & 1.90 & $3.55\times10^{16}$ & 1.71  \\
$1\times10^{-6}$ & $4.68\times10^{16}$ & 2.09 & $4.88\times10^{16}$ & 1.88 \\
$2\times10^{-6}$& $6.68\times10^{16}$ & 2.29 & $6.94\times10^{16}$ & 2.05  \\
$5\times10^{-6}$ & $1.09\times10^{17}$ & 2.58 & $1.15\times10^{17}$ & 2.29  \\
$1\times10^{-5}$ & $1.58\times10^{17}$ & 2.83 & $1.72\times10^{17}$ & 2.47 \\
$2\times10^{-5}$ & $2.29\times10^{17}$ & 3.05 & $2.61\times10^{17}$ & 2.66  \\
$5\times10^{-5}$ & $3.81\times10^{17}$ & 3.27 & $4.63\times10^{17}$ & 2.89  \\
$1\times10^{-4}$ & $5.74\times10^{17}$ & 3.44 & $7.26\times10^{17}$ & 3.07  \\
\hline
\end{tabular}
\label{Fittable}
\end{table}

%
\subsection{Predicting the CO line fluxes}\label{RT}

We have tested the effect of CO abundance profiles derived in this work and by MGH88 on the CO line intensities for two models in Table \ref{Fittable}.
We use a non-local thermodynamic equilibrium (non-LTE) RT code based on the Monte Carlo programme (mcp) \citep[see e.g.][]{Schoier01} for the excitation analysis.

Model 1. The first model is the reference model with its CSE properties are given in Table \ref{SModel}. We assumed that the reference star is at a distance of 1 kpc in order to cover the entire envelope with a 12-m telescope beam.
The fitting parameters of the CO envelope size are $r_{1/2}=2.02\times 10^{17}$ cm and $\alpha=3.10$ from this work and $r_{1/2}=2.35\times 10^{17}$ cm and $\alpha=2.79$ from MGH88.
Figure \ref{RT1} presents the results of RT modelling for CO(1-0, 2-1, 3-2) transitions for this model. 
The $14\%$ difference in $r_{1/2}$ size only becomes visible in $J=1-0$ spectrum in this case. We find a $5\%$ difference in the total intensity of $J=1-0$ transition which is less than the commonly assumed $20\%$ uncertainty of single-dish observations.
However, comparing the integrated intensity at radial offset points from the central star shows bigger discrepancy between two models. The differences in the integrated intensity at distance $22'' = 3.2 \times 10^{17}$ cm reaches to $34\%$. This indicates that CO high-resolution observations which provide the integrated intensity at radial offset points are more powerful to constrain photodissociation models.

Model 2. We selected the model with the biggest discrepancy with MGH88 work. This model has $\dot{M}=10^{-4} \: [M_{\odot} \rm yr^{-1}]$ and $V_{\rm exp}=7.5 \: [\rm km \: s^{-1}]$. The star is assumed to be at distance of 3 kpc from the earth to cover the entire envelope with a 12-m telescope beam.
The fitting parameters of the CO envelope size are $r_{1/2}=9.91\times 10^{17}$ cm and $\alpha=3.80$ from this work and $r_{1/2}=1.64\times 10^{18}$ cm and $\alpha=3.71$ from MGH88.
Figure \ref{RT2} shows the results of the CO RT modelling for these two models. The clearest difference is again apparent for CO(1-0), where the emission is substantially different in the velocity range [-10,-5] [km s$^{-1}$]. In the blue model, the emission is entirely self-absorbed. This can be explained by the absence of a significant amount of cold CO gas in the red model, which is characterised by a smaller $r_{1/2}$ than the blue model.
For this model, the integrated intensities at radial offset points of the red model (this work) are higher due to the absorption in the blue model (MGH88).

\begin{figure}[t]
  \centering
  \includegraphics[width=90mm]{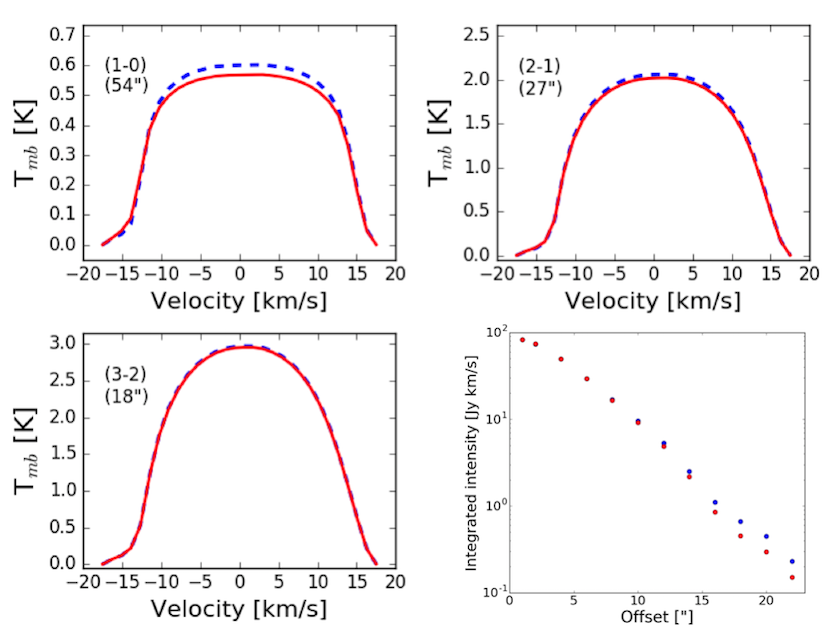}
  \caption[]{\label{RT1}
Results of CO RT modelling for model 1 (the reference model) with the CO abundance distribution from this work (red) and MGH88 (blue). The transitions and the beam size are marked in each panel. The bottom right panel compares the radial distribution of the CO(1-0) integrated intensities at radial offset points.}
\label{RT1}  
\end{figure}

\begin{figure}[t]
  \centering
  \includegraphics[width=95mm]{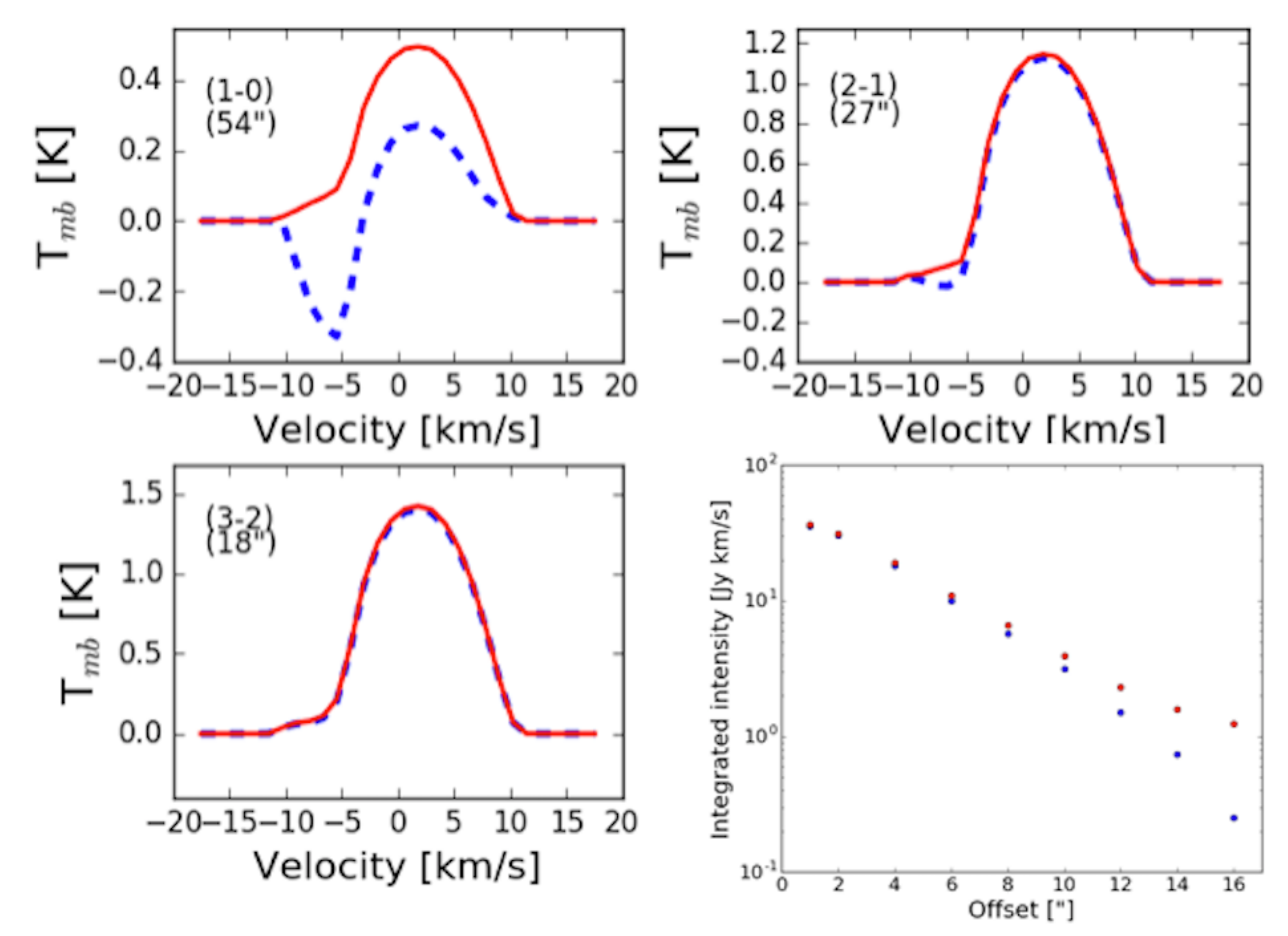}
  \caption[]{\label{RT2}
Results of CO RT modelling for model 2 with extreme discrepancy between the CO abundance distribution from this work (red) and MGH88 (blue). The transitions and the beam size are marked in each panel. The bottom right panel compares the radial distribution of the CO(1-0) integrated intensities at radial offset points.}
\label{RT2}  
\end{figure}

\section{Summary}\label{Summary}

We have presented detailed calculations of the CO photodissociation rate in a spherically symmetric CSE which is expanding with a constant velocity. The standard \cite{Draine78} radiation field is assumed to penetrate into the CSE from all directions. We used the latest CO spectroscopic data to calculate the shielding functions.
We examined the impact of variation of five primary important factors $\dot{M}$, $V_{\rm exp}$, $f_{\rm CO}$, $T_{\rm ex}$(CO), and the strength of the ISRF $\chi$ on the CO abundance distributions.
The effect of varying parameters on the CO envelope size is more prominent for either low-mass loss stars or the ones with low initial CO abundance. This can be explained by lower shielding efficiency. 

Assuming the same ISRF and CSE properties, our derived CO envelope size is smaller than the commonly used radius presented by MGH88. 
We show that having the same amount of effective CO with a different set of physical parameters does not necessarily give the same CO abundance distribution. 
High-resolution ALMA observations, for example the DEATHSTAR\footnote{http://www.astro.uu.se/deathstar/index.html} project (Ramstedt et al. In prep), can, together with our new formalism of determination of the CO envelope size, be used to decrease the uncertainty in mass-loss determinations.

Although we listed two fitting parameters of the CO abundance distribution for a large grid of models, it is recommended to run models for individual stars separately considering their individual physical parameters. 
Moreover, we strongly recommend running optimised models in case there are clear indications for a locally weak or strong ISRF based on other observations.
\begin{acknowledgements}

This work was supported by ERC consolidator grant 614264. EDB acknowledges financial support from the Swedish National Space Agency. We thank John Black for fruitful discussions on the detailed calculations of photodissociation rates, providing the H$_2$ data list, and commenting on the manuscript, Tom Millar for several discussions and explanations of the original circumstellar chemistry code and commenting on the manuscript, Theo Khouri for useful discussions, Evelyne Roueff and Franck Le Petit for useful discussions on PDR models. We are grateful to the anonymous referee for their constructive feedback on the manuscript.

\end{acknowledgements}




\bibliographystyle{aa} 
\bibliography{refrences} 


\begin{appendix} 

\section{CO shielding functions}\label{COshieldingfunctions}

\subsection{Unshielded CO photodissociation rate}\label{The unshielded ph rate}

The contribution of each individual transition to the unshielded photodissociation rate at the outer radius can be calculated using \citep{vanDishoeck88}:
\begin{equation}
k_i^0 = \frac{\pi e^2}{m_e c^2} \: f_i \: \eta_i \:  x_l \:  \lambda_i^2 \: I_{\rm ISRF}(\lambda_i) \:\:\:  [s^{-1}],
\label{ki0}
\end{equation}
where $f$ is the absorption oscillator strength that expresses the probability of absorption of electromagnetic radiation in transitions between energy levels of a molecule, $\eta$ is the probability for dissociation of the upper level, $x_l$ is the fractional populations of the lower level and $I_{\rm ISRF}$ is the mean intensity of the interstellar radiation field in unit [photons cm$^{-2}$ s$^{-1}$ \r{A}$^{-1}$]. The constant factor $ {\pi e^2 / m_e c^2} $ takes the value 8.85 $\times$ 10$^{-21}$ if $\lambda$ is in \r{A}. We have assumed $x_l$ has a Boltzmann distribution profile as follow:

\begin{equation}
x_{l} = \frac{g_l \: \exp(-E_{l}/kT_{\rm ex})}{\sum_{l=0}^{9} \: g_l \: \exp(-E_{l}/kT_{\rm ex})} ,
\label{xl}
\end{equation}
where $g_l$ and $E_{\rm l}$ are the degeneracy and energy of the lower level. 
Since in our model $T_{\mathrm ex}$(CO) =  $T_{\rm kin}$, to derive the fractional population of lower levels at the edge of the CSE, we considered $T_{\mathrm ex}$ = 10 K which is the minimum acceptable temperature in our chemical network.

We considered the standard Draine radiation field \cite{Draine78} as the ISRF  which has the form:
\begin{equation}
I_{\rm ISRF}(\lambda) = \left(\frac{6.36 \times 10^7}{\lambda^4} - \frac{1.0237 \times 10^{11}}{\lambda^5} + \frac{4.0812 \times 10^{13}}{\lambda^6} \right),
\label{IISRF}
\end{equation}
here $I$ is in unit [ergs cm$^{-2} s^{-1} \r{A}^{-1}$]. We multiplied $I$ by a factor $(5.03 \times 10^7 \times \lambda)$ to convert the intensity unit to [photons cm$^{-2}$ s$^{-1}$ \r{A}$^{-1}$] which is needed in Eq. \ref{ki0}.

\subsection{CO self-shielding}\label{CO self-shielding}

The CO self-shielding in an expanding spherical envelope was approximated by MJ83 to be: 

\begin{equation}
\beta_i (r) = \frac{1-\exp(-1.5 \: \tau_{CO} (\nu_i,r))}{1.5 \: \tau_{CO} (\nu_i,r)},
\label{betai}
\end{equation}
where $\beta_i(r)$ indicates the escape probability for a photon which is generated at radius $r$. 
When the absorption occurs primarily in the Doppler core of the line and by assuming that the CSE expansion velocity $V_{\rm exp}$ dominates the line width, the optical depth at the centre of each dissociating line $\nu_i$ can be written:

\begin{equation}
\tau_{CO}(\nu_i, r) = \frac{\pi e^2}{m_e c}  \: N_{CO}(r) \: x_l(r) \: f_{ij} \: \lambda_i \: \frac{1}{V_{\rm exp}} + \sum_j \tau_{CO}(\nu_j,r) \: ,
\label{tau(CO)}
\end{equation}
where $N_{\rm CO}$ is a column density integrated from outside to r. The first term counts the CO optical depth at the centre of each dissociating line $i$ and the second term counts the effect of all blended lines $j$  if $\lambda_i - \lambda_j < \Delta\lambda_j$. Here, $\Delta\lambda_j$ is the CO line broadening. 
In the first term $\pi e^2 / m_e c = 0.0265$ and $\lambda$ is in cm and $V_{\rm  exp}$ is in cm s$^{-1}$. The fractional population of the lower level $x_l$ varies by the radius here.

To derive $\tau_{CO}(\nu_j,r)$ we have to estimate the line width of each dissociating line. Since the effect of the expansion velocity is equivalent for all lines, we need to only consider the thermal and natural broadenings.

\paragraph{\bf Thermal broadening}
The thermal or Doppler broadening due to the random thermal motions of molecules has a gaussian profile as follows:
\begin{equation}
\phi_\nu = \frac{1}{\sqrt\pi} \frac{1}{\Delta\nu_{th}} \rm exp\left\{-\Big(\frac{\nu - \nu_0}{\Delta\nu_{th}}\Big)^2\right\},
\label{phinuT}
\end{equation}
where the line of sight thermal width is defined as
\begin{equation}
v_{\rm th}  \equiv \sqrt{\frac{2k T_k}{m_{\rm CO}}},
\label{vth}
\end{equation}
and in frequency units:
\begin{equation}
  \Delta\nu_{\rm th} = \frac{\nu_0}{c} \sqrt{\frac{2k T_k}{m_{\rm CO}}}.
 \label{Deltanu} 
\end{equation}
The peak value of $\phi(\nu = \nu_0)$ is given by
\begin{equation}
  \phi_0 = \frac{1}{\sqrt \pi} \: \frac{1}{\Delta\nu_{\rm th}},
  \label{phi0}
\end{equation}
and the full-width at half maximum (FWHM):
\begin{equation}
  \Delta v_{1/2} = 1.665 \:  v_{\rm th},
  \label{Deltanu12}
\end{equation}
so the CO thermal broadening across the temperature range 10-2000 K is equivalent to 0.13 - 1.8 km s$^{-1}$. These give the marginal line broadenings of $\Delta \lambda_{1/2} = (\Delta v_{1/2} / c) \: \lambda \sim 0.4 \times 10 ^{-3} -0.6 \times 10^{-2} \r{A}$ at the wavelength of $1000 \: \r{A}$. Thus, we can ignore the thermal line broadening of CO.

\paragraph{\bf Natural broadening} 
The natural line broadening resulting from the Heisenberg uncertainty principle gives a Lorentzian profile:
\begin{equation}
  \phi_\nu = \frac{\Gamma}{4 \pi^2 (\nu - \nu_0)^2 + (\Gamma/2)^2},
  \label{phinuN}
\end{equation}
where $\Gamma$ is the quantum-mechanical damping constant and represents the total radiative decay probability or the inverse radiative lifetime of the upper level in s$^{-1}$. 
The peak value of $\phi (\nu=\nu_0)$ at the line centre is given by
\begin{equation}
   \phi_0 = \frac{4}{\Gamma},
 \end{equation}
 and the FWHM in frequency units will be: 
\begin{equation}
  \Delta \nu_{1/2} = \frac{\Gamma}{2\pi}.
\end{equation}

CO has $\Gamma \sim10^8 - 10^{13}$ s$^{-1}$ which leads to the line broadening $\Delta \lambda_{1/2} = (\lambda^2/C) \Delta\nu_{1/2} \sim (0.5 \times10^{-5} - 0.5) \: \r{A}$. Thus, for the lines with large $\Gamma$ the natural broadenings are significant.

As we now have the precise line positions and line widths of each individual line, we can estimate the number and contribution of blended lines in the opacity of each dissociating lines. We assumed that if $\lambda_i - \lambda_j < \Delta\lambda_j$, the j line is be considered in the second term of Eq. \ref{tau(CO)}.

\subsection{Mutual-shielding by H$_2$ and dust }                   

The shielding of CO by molecular hydrogen, H$_2$, and dust in an expanding CSE is numerically approximated by MJ83 to be:

\begin{equation}
\gamma_i (r) = \exp \Big(-\alpha \: \big( \tau_{\rm dust}(r)  + \tau_{\rm H_2}(\nu_i,r) \big)^b\Big),
\label{gammai}
\end{equation}
where $\alpha = 1.644$, $b=0.86$. $\tau_{\rm H_2}$ is the H$_2$ opacity at each CO dissociating line $i$ and can be calculated as:

\begin{equation}
\tau_{H_2}(\nu_i,r) =  \frac{\pi e^2}{m_e c} f_m \phi_{\nu_i} N_{\rm H_2},
\label{tauH21}
\end{equation}
here $\phi_{\nu_i}$ is the H$_2$ line shape at the CO line $i$. 
Since the H$_2$ lines become optically thick, the absorption occurs in the radiative wings of the Lyman and Werner transitions. 
Thus, CO mutual-shielding by H$_2$ mostly occurs at the Lorentzian damping wings of the H$_2$ line profile (MJ83, MGH88) which is presented in Eq. \ref{phinuN}. 
This gives the H$_2$ opacity at each CO dissociating line $i$:

\begin{equation}
\tau_{H_2}(\nu_i,r) = \sum_m \frac{\pi e^2}{m_e c} f_{m} N_{\rm H_2}  \frac{\Gamma_m}{4 \pi^2 (\nu_m - \nu_i)^2 + (\Gamma_m/2)^2},
\label{tauH22}
\end{equation}
where $m$ sums over all H$_2$ dissociating lines.

In Eq. \ref{gammai}, we assumed that the dust absorption is independent of the wavelength in the spectral region of interest. We also ignored dust scattering and assume that the dust absorption dominates the dust extinction.

Therefore, we considered a constant dust extinction at $1000 \: \r{A}$, as MJ83, to be:

\begin{equation}
\tau_{dust} (r, 1000 \r{A}) = \frac{4.65 \times 2 \times N_{\rm H_2}(r)}{1.87 \times 10^{21}},
\label{taudust}
\end{equation}
where $N_{\rm H_2}$ is the H$_2$ column density to infinity.

\clearpage
\onecolumn

\section{Grid models} \label{The grid models}

Table \ref{FullG} presents the two fitting parameters of the CO envelope size for the full grid of models.

\begin{longtable}{c|cc|cc|ccc}
\caption{Fitting parameters $\alpha$ and $r_{1/2}$ which approximate the CO envelope size for all models.}\\
\hline
\multicolumn{7}{c}{$f_{\rm CO/H_2}= 1\times 10^{-4}$} \\
 \hline  
$\dot{M} \: [M_{\sun} \rm yr^{-1}]$  &   $V_{\rm exp}= 7.5 \: [\rm km \: s^{-1}]$ &   &   $V_{\rm exp}= 15 \: [\rm km \: s^{-1}]$   &    & $V_{\rm exp}= 30 \: [\rm km\: s^{-1}]$ &  \\  
\hline
 & $\alpha$  &   $r_{1/2}$ [cm] &  $\alpha$ &   $r_{1/2}$ [cm] &  $\alpha$ & $r_{1/2}$ [cm]   \\
\hline     
1$\times 10^{-8}$ & 1.48 & 4.41$\times 10^{15}$ & 1.21 & 5.91$\times 10^{15}$ & 1.10 & 9.71$\times 10^{15}$\\
2$\times 10^{-8}$ & 1.65 & 5.51$\times 10^{15}$ & 1.30 & 6.75$\times 10^{15}$ & 1.14 & 1.03$\times 10^{16}$\\
5$\times 10^{-8}$ & 1.95 & 7.81$\times 10^{15}$& 1.45 & 8.48$\times 10^{15}$ & 1.21 & 1.17$\times 10^{16}$\\
1$\times 10^{-7}$ & 2.20 & 1.06$\times 10^{16}$& 1.61 & 1.05$\times 10^{16}$  & 1.29 & 1.32$\times 10^{16}$\\
2$\times 10^{-7}$ & 2.47 & 1.48$\times 10^{16}$& 1.82 & 1.37$\times 10^{16}$ & 1.40 & 1.56$\times 10^{16}$\\
5$\times 10^{-7}$ & 2.79 & 2.38$\times 10^{16}$& 2.16 & 2.07$\times 10^{16}$ & 1.60 & 2.06$\times 10^{16}$\\
1$\times 10^{-6}$ & 3.03 & 3.47$\times 10^{16}$& 2.46 & 2.94$\times 10^{16}$ & 1.81 & 2.68$\times 10^{16}$\\
2$\times 10^{-6}$ & 3.27 & 5.09$\times 10^{16}$& 2.76 & 4.26$\times 10^{16}$ & 2.07 & 3.67$\times 10^{16}$\\
5$\times 10^{-6}$ & 3.62 & 8.57$\times 10^{16}$& 3.15 & 7.13$\times 10^{16}$ & 2.48 & 5.85$\times 10^{16}$\\
1$\times 10^{-5}$ & 3.96 & 1.28$\times 10^{17}$& 3.42 & 1.06$\times 10^{17}$ & 2.81 & 8.58$\times 10^{16}$\\
2$\times 10^{-5}$ & 4.35 & 1.95$\times 10^{17}$& 3.72 & 1.60$\times 10^{17}$ & 3.14 & 1.27$\times 10^{17}$\\
5$\times 10^{-5}$ & 4.87 & 3.56$\times 10^{17}$& 4.08 & 2.80$\times 10^{17}$ & 3.59 & 2.19$\times 10^{17}$\\
1$\times 10^{-4}$ & 5.34 & 5.76$\times 10^{17}$& 4.37 & 4.40$\times 10^{17}$ & 3.88 & 3.37$\times 10^{17}$\\
\hline
\hline
\multicolumn{7}{c}{$f_{\rm CO/H_2}$ = $2\times 10^{-4} \:\:(\rm M-type \:AGB)$}\\
 \hline  
 \hline
1$\times 10^{-8}$ & 1.58 & 4.92$\times 10^{15}$& 1.26 & 6.20$\times 10^{15}$ & 1.11 &9.84$\times 10^{15}$\\
2$\times 10^{-8}$ & 1.77 & 6.30$\times 10^{15}$& 1.36 & 7.21$\times 10^{15}$ & 1.15 &1.06$\times 10^{16}$\\
5$\times 10^{-8}$ & 2.06 & 9.25$\times 10^{15}$& 1.54 & 9.36$\times 10^{15}$  & 1.24 &1.21$\times 10^{16}$\\
1$\times 10^{-7}$ & 2.27 & 1.28$\times 10^{16}$& 1.72 & 1.19$\times 10^{16}$ & 1.34 &1.40$\times 10^{16}$\\
2$\times 10^{-7}$ & 2.46 & 1.81$\times 10^{16}$ & 1.93 & 1.60$\times 10^{16}$  & 1.46 &1.69$\times 10^{16}$\\
5$\times 10^{-7}$ & 2.69 & 2.94$\times 10^{16}$& 2.24 & 2.48$\times 10^{16}$ & 1.69 &2.31$\times 10^{16}$\\
1$\times 10^{-6}$ & 2.85 & 4.29$\times 10^{16}$& 2.49 & 3.55$\times 10^{16}$ & 1.91 &3.10$\times 10^{16}$\\
2$\times 10^{-6}$ & 3.02 & 6.30$\times 10^{16}$& 2.72 & 5.17$\times 10^{16}$  & 2.17 &4.32$\times 10^{16}$\\
5$\times 10^{-6}$ & 3.29 & 1.05$\times 10^{17}$& 3.01 & 8.65$\times 10^{16}$ & 2.53 &6.97$\times 10^{16}$\\
1$\times 10^{-5}$ & 3.58 & 1.55$\times 10^{17}$& 3.25 & 1.28$\times 10^{17}$  & 2.81 &1.02$\times 10^{17}$\\
2$\times 10^{-5}$ & 3.86 & 2.33$\times 10^{17}$& 3.50 & 1.90$\times 10^{17}$ & 3.11 &1.51$\times 10^{17}$\\
5$\times 10^{-5}$ & 4.28 & 4.19$\times 10^{17}$& 3.76 & 3.30$\times 10^{17}$ & 3.46 &2.57$\times 10^{17}$\\
1$\times 10^{-4}$ & 4.66 & 6.70$\times 10^{17}$ & 4.00 & 5.14$\times 10^{17}$ & 3.69 &3.91$\times 10^{17}$\\
\hline
\hline
\multicolumn{7}{c}{$f_{\rm CO/H_2}$ = $3\times 10^{-4}$}\\
 \hline  
 \hline
1$\times 10^{-8}$ & 1.65 & 5.34$\times 10^{15}$& 1.29 & 6.45$\times 10^{15}$ & 1.12 &9.96$\times 10^{15}$\\
2$\times 10^{-8}$ & 1.84 & 6.96$\times 10^{15}$& 1.40 & 7.61$\times 10^{15}$ & 1.17 &1.08$\times 10^{16}$\\
5$\times 10^{-8}$ & 2.10 & 1.04$\times 10^{16}$ & 1.60 & 1.01$\times 10^{16}$ & 1.27 &1.25$\times 10^{16}$\\
1$\times 10^{-7}$ & 2.27 & 1.45$\times 10^{16}$& 1.78 & 1.31$\times 10^{16}$  & 1.37 &1.47$\times 10^{16}$\\
2$\times 10^{-7}$ & 2.43 & 2.08$\times 10^{16}$& 1.99 & 1.79$\times 10^{16}$ & 1.51 &1.80$\times 10^{16}$\\
5$\times 10^{-7}$ & 2.63 & 3.40$\times 10^{16}$& 2.27 & 2.81$\times 10^{16}$ & 1.75 &2.53$\times 10^{16}$\\
1$\times 10^{-6}$ & 2.77 & 4.97$\times 10^{16}$& 2.48 & 4.04$\times 10^{16}$ & 1.97 &3.44$\times 10^{16}$\\
2$\times 10^{-6}$ & 2.91 & 7.29$\times 10^{16}$& 2.69 & 5.90$\times 10^{16}$ & 2.22 &4.83$\times 10^{16}$\\
5$\times 10^{-6}$ & 3.16 & 1.20$\times 10^{17}$& 2.95 & 9.85$\times 10^{16}$ & 2.55 &7.84$\times 10^{16}$\\
1$\times 10^{-5}$ & 3.41 & 1.76$\times 10^{17}$& 3.18 & 1.45$\times 10^{17}$ & 2.81 &1.14$\times 10^{17}$\\
2$\times 10^{-5}$ & 3.64 & 2.63$\times 10^{17}$& 3.39 & 2.14$\times 10^{17}$ & 3.08 &1.69$\times 10^{17}$\\
5$\times 10^{-5}$ & 4.01 & 4.68$\times 10^{17}$& 3.62 & 3.69$\times 10^{17}$ & 3.40 &2.85$\times 10^{17}$\\
1$\times 10^{-4}$ & 4.35 & 7.42$\times 10^{17}$& 3.84 & 5.72$\times 10^{17}$ & 3.60 &4.33$\times 10^{17}$\\
\hline
\hline
\multicolumn{7}{c}{$f_{\rm CO/H_2}$ = $4\times 10^{-4}$}\\
 \hline  
 \hline
1$\times 10^{-8}$ & 1.70 & 5.71$\times 10^{15}$& 1.32 & 6.67$\times 10^{15}$ & 1.13 &1.00$\times 10^{16}$\\
2$\times 10^{-8}$ & 1.88 & 7.53$\times 10^{15}$& 1.44 & 7.97$\times 10^{15}$ & 1.19 &1.10$\times 10^{16}$\\
5$\times 10^{-8}$ & 2.12 & 1.14$\times 10^{16}$& 1.64 & 1.07$\times 10^{16}$ & 1.29 &1.29$\times 10^{16}$\\
1$\times 10^{-7}$ & 2.27 & 1.61$\times 10^{16}$& 1.82 & 1.42$\times 10^{16}$ & 1.40 &1.53$\times 10^{16}$\\
2$\times 10^{-7}$ & 2.42 & 2.31$\times 10^{16}$& 2.02 & 1.95$\times 10^{16}$ & 1.55 &1.90$\times 10^{16}$\\
5$\times 10^{-7}$ & 2.59 & 3.80$\times 10^{16}$ & 2.28 & 3.09$\times 10^{16}$  & 1.80 &2.72$\times 10^{16}$\\
1$\times 10^{-6}$ & 2.72 & 5.56$\times 10^{16}$& 2.48 & 4.46$\times 10^{16}$  & 2.01 &3.74$\times 10^{16}$\\
2$\times 10^{-6}$ & 2.86 & 8.14$\times 10^{16}$& 2.66 & 6.53$\times 10^{16}$ & 2.24 &5.28$\times 10^{16}$\\
5$\times 10^{-6}$ & 3.10 & 1.33$\times 10^{17}$& 2.93 & 1.08$\times 10^{17}$ & 2.55 &8.58$\times 10^{16}$\\
1$\times 10^{-5}$ & 3.32 & 1.94$\times 10^{17}$& 3.15 & 1.59$\times 10^{17}$ & 2.81 &1.25$\times 10^{17}$\\
2$\times 10^{-5}$ & 3.51 & 2.89$\times 10^{17}$& 3.33 & 2.34$\times 10^{17}$ & 3.07 &1.84$\times 10^{17}$\\
5$\times 10^{-5}$ & 3.84 & 5.10$\times 10^{17}$& 3.54 & 4.02$\times 10^{17}$  & 3.35 &3.09$\times 10^{17}$\\
1$\times 10^{-4}$ & 4.16 & 8.04$\times 10^{17}$& 3.74 & 6.20$\times 10^{17}$ & 3.55 &4.68$\times 10^{17}$\\
\hline
\hline
\multicolumn{7}{c}{$f_{\rm CO/H_2}$ = $5\times 10^{-4}$}\\
 \hline  
 \hline
1$\times 10^{-8}$ & 1.73 & 6.05$\times 10^{15}$& 1.34 & 6.88$\times 10^{15}$ & 1.14 &1.01$\times 10^{16}$\\
2$\times 10^{-8}$ & 1.91 & 8.05$\times 10^{15}$& 1.47 & 8.30$\times 10^{15}$ & 1.20 &1.11$\times 10^{16}$\\
5$\times 10^{-8}$ & 2.13 & 1.23$\times 10^{16}$& 1.67 & 1.13$\times 10^{16}$ & 1.31 &1.33$\times 10^{16}$\\
1$\times 10^{-7}$ & 2.27 & 1.74$\times 10^{16}$& 1.86 & 1.52$\times 10^{16}$ & 1.43 &1.59$\times 10^{16}$\\
2$\times 10^{-7}$ & 2.40 & 2.52$\times 10^{16}$& 2.04 & 2.10$\times 10^{16}$ & 1.58 &2.00$\times 10^{16}$\\
5$\times 10^{-7}$ & 2.57 & 4.16$\times 10^{16}$& 2.29 & 3.34$\times 10^{16}$ & 1.83 &2.90$\times 10^{16}$\\
1$\times 10^{-6}$ & 2.69 & 6.10$\times 10^{16}$& 2.47 & 4.84$\times 10^{16}$ & 2.04 &4.00$\times 10^{16}$\\
2$\times 10^{-6}$ & 2.83 & 8.90$\times 10^{16}$& 2.65 & 7.09$\times 10^{16}$ & 2.26 &5.67$\times 10^{16}$\\
5$\times 10^{-6}$ & 3.07 & 1.45$\times 10^{17}$& 2.91 & 1.17$\times 10^{17}$ & 2.56 &9.24$\times 10^{16}$\\
1$\times 10^{-5}$ & 3.25 & 2.10$\times 10^{17}$& 3.13 & 1.72$\times 10^{17}$ & 2.81 &1.35$\times 10^{17}$\\
2$\times 10^{-5}$ & 3.42 & 3.11$\times 10^{17}$& 3.29 & 2.51$\times 10^{17}$ & 3.07 &1.97$\times 10^{17}$\\
5$\times 10^{-5}$ & 3.73 & 5.47$\times 10^{17}$ & 3.48 & 4.30$\times 10^{17}$  & 3.32 &3.30$\times 10^{17}$\\
1$\times 10^{-4}$ & 4.03 & 8.58$\times 10^{17}$& 3.68 & 6.63$\times 10^{17}$ & 3.51 &4.99$\times 10^{17}$\\
\hline
\hline
\multicolumn{7}{c}{$f_{\rm CO/H_2}$ = $6\times 10^{-4} \:\:(\rm S-type \:AGB)$} \\
 \hline  
 \hline
1$\times 10^{-8}$ & 1.76 & 6.36$\times 10^{15}$& 1.36 & 7.08$\times 10^{15}$ & 1.15 &1.03$\times 10^{16}$\\
2$\times 10^{-8}$ & 1.93 & 8.53$\times 10^{15}$& 1.49 & 8.61$\times 10^{15}$ & 1.21 &1.13$\times 10^{16}$\\
5$\times 10^{-8}$ & 2.14 & 1.31$\times 10^{16}$& 1.70 & 1.19$\times 10^{16}$ & 1.33 &1.36$\times 10^{16}$\\
1$\times 10^{-7}$ & 2.27 & 1.87$\times 10^{16}$& 1.88 & 1.60$\times 10^{16}$ & 1.45 &1.65$\times 10^{16}$\\
2$\times 10^{-7}$ & 2.40 & 2.71$\times 10^{16}$& 2.06 & 2.23$\times 10^{16}$ & 1.61 &2.09$\times 10^{16}$\\
5$\times 10^{-7}$ & 2.56 & 4.49$\times 10^{16}$& 2.30 & 3.57$\times 10^{16}$ & 1.86 &3.06$\times 10^{16}$\\
1$\times 10^{-6}$ & 2.67 & 6.59$\times 10^{16}$& 2.47 & 5.18$\times 10^{16}$ & 2.06 &4.25$\times 10^{16}$\\
2$\times 10^{-6}$ & 2.81 & 9.60$\times 10^{16}$& 2.64 & 7.60$\times 10^{16}$ & 2.27 &6.04$\times 10^{16}$\\
5$\times 10^{-6}$ & 3.05 & 1.55$\times 10^{17}$& 2.91 & 1.26$\times 10^{17}$ & 2.56 &9.84$\times 10^{16}$\\
1$\times 10^{-5}$ & 3.21 & 2.24$\times 10^{17}$& 3.12 & 1.83$\times 10^{17}$ & 2.81 &1.43$\times 10^{17}$\\
2$\times 10^{-5}$ & 3.36 & 3.32$\times 10^{17}$& 3.26 & 2.67$\times 10^{17}$ & 3.06 &2.08$\times 10^{17}$\\
5$\times 10^{-5}$ & 3.66 & 5.80$\times 10^{17}$& 3.44 & 4.56$\times 10^{17}$ & 3.30 &3.48$\times 10^{17}$\\
1$\times 10^{-4}$ & 3.93 & 9.06$\times 10^{17}$& 3.62 & 7.00$\times 10^{17}$ & 3.48 &5.26$\times 10^{17}$\\
\hline
\hline
\multicolumn{7}{c}{$f_{\rm CO/H_2}$ = $7\times 10^{-4}$}\\
 \hline  
 \hline
1$\times 10^{-8}$ & 1.78 & 6.64$\times 10^{15}$ & 1.38 & 7.26$\times 10^{15}$ & 1.16 &1.04$\times 10^{16}$\\
2$\times 10^{-8}$ & 1.95 & 8.97$\times 10^{15}$& 1.51 & 8.90$\times 10^{15}$ & 1.22 &1.15$\times 10^{16}$\\
5$\times 10^{-8}$ & 2.14 & 1.39$\times 10^{16}$& 1.72 & 1.24$\times 10^{16}$ & 1.35 &1.40$\times 10^{16}$\\
1$\times 10^{-7}$ & 2.27 & 1.99$\times 10^{16}$& 1.90 & 1.69$\times 10^{16}$ & 1.48 &1.70$\times 10^{16}$\\
2$\times 10^{-7}$ & 2.39 & 2.89$\times 10^{16}$& 2.07 & 2.36$\times 10^{16}$ & 1.63 &2.17$\times 10^{16}$\\
5$\times 10^{-7}$ & 2.55 & 4.80$\times 10^{16}$& 2.30 & 3.78$\times 10^{16}$ & 1.88 &3.21$\times 10^{16}$\\
1$\times 10^{-6}$ & 2.66 & 7.04$\times 10^{16}$& 2.47 & 5.50$\times 10^{16}$ & 2.08 &4.47$\times 10^{16}$\\
2$\times 10^{-6}$ & 2.80 & 1.02$\times 10^{17}$ & 2.64 & 8.08$\times 10^{16}$ & 2.28 &6.37$\times 10^{16}$\\
5$\times 10^{-6}$ & 3.03 & 1.64$\times 10^{17}$& 2.91 & 1.33$\times 10^{17}$ & 2.57 &1.03$\times 10^{17}$\\
1$\times 10^{-5}$ & 3.17 & 2.37$\times 10^{17}$& 3.11 & 1.93$\times 10^{17}$ & 2.82 &1.51$\times 10^{17}$\\
2$\times 10^{-5}$ & 3.32 & 3.50$\times 10^{17}$& 3.24 & 2.81$\times 10^{17}$ & 3.06 &2.19$\times 10^{17}$\\
5$\times 10^{-5}$ & 3.59 & 6.10$\times 10^{17}$& 3.41 & 4.79$\times 10^{17}$ & 3.28 &3.65$\times 10^{17}$\\
1$\times 10^{-4}$ & 3.87 & 9.50$\times 10^{17}$& 3.59 & 7.35$\times 10^{17}$ & 3.46 &5.51$\times 10^{17}$\\
\hline
\hline
\multicolumn{7}{c}{$f_{\rm CO/H_2}$ = $8\times 10^{-4}$}\\
 \hline  
 \hline
1$\times 10^{-8}$ & 1.80 & 6.92$\times 10^{15}$& 1.40 & 7.44$\times 10^{15}$ & 1.16 &1.04$\times 10^{16}$\\
2$\times 10^{-8}$ & 1.96 & 9.38$\times 10^{15}$& 1.53 & 9.18$\times 10^{15}$ & 1.23 &1.17$\times 10^{16}$\\
5$\times 10^{-8}$ & 2.14 & 1.46$\times 10^{16}$& 1.74 & 1.29$\times 10^{16}$  & 1.36 &1.43$\times 10^{16}$\\
1$\times 10^{-7}$ & 2.27 & 2.10$\times 10^{16}$& 1.91 & 1.76$\times 10^{16}$ & 1.49 &1.75$\times 10^{16}$\\
2$\times 10^{-7}$ & 2.39 & 3.06$\times 10^{16}$& 2.08 & 2.47$\times 10^{16}$ & 1.65 &2.25$\times 10^{16}$\\
5$\times 10^{-7}$ & 2.54 & 5.09$\times 10^{16}$& 2.30 & 3.98$\times 10^{16}$ & 1.90 &3.35$\times 10^{16}$\\
1$\times 10^{-6}$ & 2.66 & 7.47$\times 10^{16}$& 2.47 & 5.81$\times 10^{16}$ & 2.09 &4.68$\times 10^{16}$\\
2$\times 10^{-6}$ & 2.80 & 1.08$\times 10^{17}$& 2.64 & 8.52$\times 10^{16}$ & 2.29 &6.68$\times 10^{16}$\\
5$\times 10^{-6}$ & 3.02 & 1.73$\times 10^{17}$& 2.91 & 1.40$\times 10^{17}$ & 2.58 &1.09$\times 10^{17}$\\
1$\times 10^{-5}$ & 3.15 & 2.49$\times 10^{17}$& 3.10 & 2.02$\times 10^{17}$ & 2.83 &1.58$\times 10^{17}$\\
2$\times 10^{-5}$ & 3.28 & 3.68$\times 10^{17}$& 3.22 & 2.95$\times 10^{17}$  & 3.05 &2.29$\times 10^{17}$\\
5$\times 10^{-5}$ & 3.55 & 6.38$\times 10^{17}$& 3.39 & 5.01$\times 10^{17}$ & 3.27 &3.81$\times 10^{17}$\\
1$\times 10^{-4}$ & 3.80 & 9.91$\times 10^{17}$ & 3.56 & 7.67$\times 10^{17}$ & 3.44 &5.74$\times 10^{17}$\\
\hline
\hline
\multicolumn{7}{c}{$f_{\rm CO/H_2}$ = $9\times 10^{-4}$}\\
 \hline  
 \hline
1$\times 10^{-8}$ & 1.82 & 7.17$\times 10^{15}$ & 1.41 & 7.60$\times 10^{15}$ & 1.17 &1.05$\times 10^{16}$\\
2$\times 10^{-8}$ & 1.97 & 9.77$\times 10^{15}$& 1.55 & 9.45$\times 10^{15}$ & 1.24 &1.18$\times 10^{16}$\\
5$\times 10^{-8}$ & 2.15 & 1.53$\times 10^{16}$& 1.76 & 1.34$\times 10^{16}$ & 1.38 &1.46$\times 10^{16}$\\
1$\times 10^{-7}$ & 2.27 & 2.20$\times 10^{16}$& 1.93 & 1.84$\times 10^{16}$ & 1.51 &1.80$\times 10^{16}$\\
2$\times 10^{-7}$ & 2.38 & 3.22$\times 10^{16}$& 2.09 & 2.58$\times 10^{16}$  & 1.67 &2.33$\times 10^{16}$\\
5$\times 10^{-7}$ & 2.53 & 5.37$\times 10^{16}$& 2.30 & 4.17$\times 10^{16}$ & 1.91 &3.48$\times 10^{16}$\\
1$\times 10^{-6}$ & 2.65 & 7.87$\times 10^{16}$& 2.47 & 6.09$\times 10^{16}$ & 2.10 &4.88$\times 10^{16}$\\
2$\times 10^{-6}$ & 2.80 & 1.13$\times 10^{17}$& 2.64 & 8.94$\times 10^{16}$ & 2.30 &6.98$\times 10^{16}$\\
5$\times 10^{-6}$ & 3.01 & 1.81$\times 10^{17}$& 2.92 & 1.46$\times 10^{17}$ & 2.59 &1.13$\times 10^{17}$\\
1$\times 10^{-5}$ & 3.12 & 2.61$\times 10^{17}$& 3.10 & 2.11$\times 10^{17}$ & 2.83 &1.64$\times 10^{17}$\\
2$\times 10^{-5}$ & 3.25 & 3.84$\times 10^{17}$& 3.20 & 3.07$\times 10^{17}$ & 3.05 &2.38$\times 10^{17}$\\
5$\times 10^{-5}$ & 3.51 & 6.64$\times 10^{17}$ & 3.37 & 5.21$\times 10^{17}$  & 3.25 &3.96$\times 10^{17}$\\
1$\times 10^{-4}$ & 3.75 & 1.02$\times 10^{18}$& 3.53 & 7.96$\times 10^{17}$ & 3.43 &5.95$\times 10^{17}$\\
\hline
\hline
\multicolumn{7}{c}{$f_{\rm CO/H_2}$ = $10\times 10^{-4}\:\: (\rm C-type \: AGB)$} \\
 \hline  
 \hline
1$\times 10^{-8}$ & 1.83 & 7.42$\times 10^{15}$& 1.43 & 7.77$\times 10^{15}$  & 1.18 &1.06$\times 10^{16}$\\
2$\times 10^{-8}$ & 1.98 & 1.01$\times 10^{16}$& 1.57 & 9.70$\times 10^{15}$ & 1.25& 1.20$\times 10^{16}$\\
5$\times 10^{-8}$ & 2.15 & 1.59$\times 10^{16}$ & 1.77 & 1.39$\times 10^{16}$ & 1.39 &1.48$\times 10^{16}$\\
1$\times 10^{-7}$ & 2.27 & 2.30$\times 10^{16}$& 1.94 & 1.91$\times 10^{16}$ & 1.53 &1.85$\times 10^{16}$\\
2$\times 10^{-7}$ & 2.38 & 3.37$\times 10^{16}$ & 2.10 & 2.69$\times 10^{16}$ & 1.69 & 2.40$\times 10^{16}$\\
5$\times 10^{-7}$ & 2.53 & 5.64$\times 10^{16}$& 2.31 & 4.35$\times 10^{16}$ & 1.92 &3.61$\times 10^{16}$\\
1$\times 10^{-6}$ & 2.65 & 8.26$\times 10^{16}$& 2.47 & 6.37$\times 10^{16}$ & 2.11 &5.07$\times 10^{16}$\\
2$\times 10^{-6}$ & 2.80 & 1.19$\times 10^{17}$& 2.65 & 9.34$\times 10^{16}$ & 2.30 &7.26$\times 10^{16}$\\
5$\times 10^{-6}$ & 3.01 & 1.89$\times 10^{17}$& 2.92 & 1.52$\times 10^{17}$ & 2.59 &1.18$\times 10^{17}$\\
1$\times 10^{-5}$ & 3.11 & 2.71$\times 10^{17}$& 3.09 & 2.19$\times 10^{17}$ & 2.84 &1.71$\times 10^{17}$\\
2$\times 10^{-5}$ & 3.23 & 3.99$\times 10^{17}$ & 3.19 & 3.19$\times 10^{17}$ & 3.04 &2.47$\times 10^{17}$\\
5$\times 10^{-5}$ & 3.44 & 6.90$\times 10^{17}$& 3.35 & 5.40$\times 10^{17}$ & 3.24 &4.09$\times 10^{17}$\\
1$\times 10^{-4}$ & 3.72 & 1.06$\times 10^{18}$& 3.52 & 8.24$\times 10^{17}$  & 3.41 &6.15$\times 10^{17}$\\
 \hline  
\label{FullG}\\
\end{longtable}

 \end{appendix}

\end{document}